\documentclass[12pt]{article}
\pdfoutput=1

\usepackage{times}
\usepackage{bm}
\usepackage{natbib}
\usepackage{amsthm}

\def\spacingset#1{\renewcommand{\baselinestretch}%
{#1}\small\normalsize} \spacingset{1}
\spacingset{1.45}

\usepackage{color}

\newcommand{\alex}[1]{{\color{black}{#1}}}

\usepackage{rotating} \usepackage{graphicx} \usepackage{subcaption}

\usepackage{float} \usepackage{bbm}

\usepackage{amsmath, amssymb} \usepackage{mathrsfs}
\usepackage{subcaption}
\usepackage[normalem]{ulem}

\usepackage{tikz}
\usepackage{tikz-3dplot}

\newtheorem{theorem}{Theorem} 
\newtheorem{lemma}{Lemma}
\newtheorem{definition}{Definition}

\usepackage{mhequ}
\newcommand{\be}{\begin{equation}\begin{aligned} }
\newcommand{\ee}{\end{aligned}\end{equation} }
\newcommand{\bb}[1]{\mathbb{#1}} \newcommand{\mc}[1]{\mathcal{#1}}
 \DeclareMathOperator{\No}{No}

 \DeclareMathOperator{\diag}{diag}

\newcommand{\dt}{\epsilon} 
\newcommand{\mass}{M} 
\newcommand{\hess}{\mathbf{H}} 
\newcommand{\e}{\ensuremath{ \mbox{\scriptsize{E}} }}

\title{\textbf{Bayesian Constraint Relaxation}}

\author{Leo L Duan, Alexander L Young, Akihiko Nishimura, David B Dunson} 

\date{}
\begin{document}

\maketitle 
\begin{abstract}
Prior information often takes  the form of parameter constraints. Bayesian
methods include such information through prior distributions having constrained
support. By using posterior sampling algorithms, one can quantify uncertainty
without relying on asymptotic approximations. However, {\em sharply}
constrained priors are (a) not necessary in some settings; and (b) tend to limit
modeling scope to a narrow set of distributions that are tractable computationally.
Inspired by the vast literature that replaces the slab-and-spike prior with a continuous approximation, we propose to replace the sharp
indicator function of the constraint with an exponential kernel, thereby creating a close-to-constrained neighborhood within the Euclidean space in which the constrained subspace is embedded.  This
kernel decays with distance from the constrained space at a rate depending on 
a relaxation hyperparameter.  By avoiding the sharp constraint, we enable use of
off-the-shelf posterior sampling algorithms, such as Hamiltonian Monte Carlo, 
facilitating automatic computation in broad models. We study the constrained and 
relaxed distributions under multiple settings, and theoretically quantify their 
differences. We illustrate the method through multiple novel modeling examples.
\end{abstract}

\section{Introduction}

It is extremely common to put constraints on parameters in statistical models. In multivariate data analysis, for example, orthogonality constraints are routinely placed on latent variables to achieve dimension reduction \citep{Cook2015}.  Other common examples
include shape constraints on functions, norm constraints on vectors, and rank constraints on matrices. There is a very rich literature on optimization subject to parameter constraints. A common approach relies on Lagrange
and Karush-Kuhn-Tucker multipliers \citep{boyd2004convex}. However,
producing a point estimate is often insufficient, as uncertainty
quantification is a key component of most statistical analyses. Conventional
large sample asymptotic theory, for example showing asymptotic normality of
statistical estimators, tends to break down in constrained inference
problems. Instead, limiting distributions may have a complex form that
needs to be re-derived for each new type of constraint and may be
intractable.

An appealing alternative is to rely on Bayesian
methods for uncertainty quantification, including the constraint through a prior distribution
having restricted support, and then applying Markov chain Monte Carlo
 to avoid the need for large sample approximations \citep{gelfand1992bayesian}. Although 
this strategy appears conceptually simple, there are three clear limitations
in practice.  First, the sharply constrained space may have smaller dimension than the Euclidean space in which it resides. This can be problematic in common statistical analysis such as model selection where a nested model in a lower dimension will always have zero posterior model probability \citep{Berger1987}. Second, when the constrained space is non-Euclidean, the choice of distributions for modeling can be quite limited. For example, under spherical constraints almost all distributions are exclusively from the von-Mises family \citep{mardia1975statistics}. Third, it is in general very difficult to develop tractable posterior sampling algorithms except in special cases.
For example, one may be forced to focus on particular forms for the prior and likelihood function to gain tractability and to develop
specially tailored algorithms on a case-by-case basis.  

Our solution is to allow deviations from the constrained space. We are largely inspired by the recent flourish of Bayesian shrinkage literature for variable selection problems \citep{George1993, armagan2013generalized, carvalho2010horseshoe}, where the key idea, roughly speaking, is to relax the exactly zero variance of the redundant parameters \citep{Mitchell1988} to a small neighborhood close to zero. We generalize this idea of a point constraint to a large set of models with equality constraints. To draw an analogy, when the intent of  constraints is to reduce the parameter variance and/or improve model interpretation, it is unnecessary to strictly uphold the sharp constraints. As an appealing alternative, the close-to-constraint posterior achieves the same goal, while being significantly more advantageous in modeling flexibility and being well-defined via standard tools of probability such as Lebesgue measure.

There is a large statistical literature on relaxation methods. For example,  \cite{srebro2005maximum} relaxes the low-rank constraint in matrix models by a continuous nuclear-norm regulariztion, \cite{Keys2016} proposes to relax the constraint to produce a surrogate function amenable to efficient optimization algorithms, and \cite{neal2011mcmc} suggests replacing an inequality constraint in Hamiltonian Monte Carlo, with a large penalty to prohibit posterior probability outside of a constrained region. Despite some similarity with the present work, the primary focus of those articles is on computation. To our best knowledge, there has been little theoretical investigation which quantifies the impact of relaxation.

The primary contribution of this article is to propose a broad class of Bayesian priors that are formally close to a constrained space and can effectively solve our problems simultaneously.  The proposed class is very broad and acts to modify an initial unconstrained prior, having an arbitrary form, to be concentrated near
the constrained space to an extent controlled by a hyperparameter.  In addition, due to the simple form and lack of any sharp parameter constraints, general off-the-shelf sampling algorithms can be applied directly.

\section{Constraint Relaxation Methodology}

\subsection{Notation and Framework}

Assume that $\theta \in \mathcal{D} \subset \mathcal{R}$ is an unknown continuous parameter, 
with $\dim(\mathcal{R})=r < \infty$.  The constrained sample space $\mathcal{D}$ is 
embedded in the $r$-dimensional Euclidean space $\mathcal{R}$. While our proposed approach is sufficiently general to handle constrained subspaces with positive or zero Lebesgue measure, the focus herein is on the zero measure case. Details regarding positive measure constrained subspaces are provided in the supplementary materials. 

The traditional Bayesian approach to including constraints requires a prior density $\pi_\mathcal{D}(\theta)$ with support on $\mc D$. The posterior density of $\theta$
given data $Y$ and $\theta \in \mathcal{D}$ is then
\begin{equation}
\pi_{\mc D}(\theta \mid  Y) 
 \propto  \pi_\mathcal{D}(\theta)\mathcal{L}(\theta; Y), \label{eq:sharp}
\end{equation}
where $\mathcal{L}(\theta; Y)$ is the likelihood function. We assume in the sequel that the restricted prior $\pi_\mathcal{D}(\theta) \propto \pi_{\mc R}(\theta)\mathbbm{1}_{\mc D}(\theta)$, with $\pi_{\mc R}(\theta)$ a distribution on ${\mc R}$ and $\mathbbm{1}_{\mc D}(\theta)$ an indicator
function that the constraint is satisfied. 

In attempting to address the issues raised in the introduction, we propose to replace (\ref{eq:sharp}) with the following constraint relaxed posterior density: 
\begin{equation}
\label{EQ:Rel_Dens_Motivation}
\tilde{\pi}_{\lambda}(\theta) \propto
\mathcal{L}(\theta; Y)  \pi_\mathcal{R}(\theta)
\exp\big(-\lambda^{-1} \|\nu_\mathcal{D}(\theta)\|\big),
\end{equation}
where we repress the conditioning on data $Y$ in $\tilde{\pi}_{\lambda}(\theta)$ for concise notation and use $\|\nu_\mathcal{D}(\theta)\|$ as
a `distance' from $\theta$ to the constrained space. 
We assume $\pi_{\mc R}(\theta)$ is proper, is supported on a set which has non-empty intersection with $\mc D$, and is absolutely continuous with respect to Lebesgue measure $\mu_\mathcal{R}$ on $\mathcal{R}$. As such, the constraint relaxed posterior $\tilde{\pi}_\lambda(\theta )$ corresponds to a coherent Bayesian probability model. 

The hyperparameter $\lambda > 0$ controls how concentrated the prior is around 
${\mc D}$ so that $\tilde{\pi}(\theta) \to 0$ as $\lambda \to 0$ for all $\theta \notin \mc D.$ 
However, for all $\lambda > 0$, $\tilde{\pi}_\lambda(\theta)$ introduces support outside
of $\mc D$, creating a relaxation of the constraint.  Both the value of $\lambda$ and the choice of $\|\nu_\mathcal{D}(\theta)\|$ are important in controlling the concentration of the prior around $\mc D$.

\subsection{Constructing $\|\nu_{\mc D}(\theta)\|$}

Many $\mu_{\mc R}$-measure zero constrained subspaces, such as the simplex or Stiefel manifold, are submanifolds of $\mc R$ which arise through equations involving $\theta$. Thus, it is natural to restrict ourselves to the setting in which $\mathcal{D}$ can be represented implicitly
as the solution set of a consistent system of equations $\{\nu_j(\theta) =
0\}_{j=1}^s$. The constraint functions, $\{\nu_j\}_{j=1}^s$, must satisfy additional assumptions as stated in Section \ref{SEC:Zero_measure_theory}. For the moment, we highlight their use in defining $\|\nu_{\mc D}(\theta)\|$.   

Given a set of constraint functions, we let $\nu_{\mc D}(\theta) =[\nu_1(\theta),\dots,\nu_s(\theta)]^T$ be a vector valued function from $\mc R$ to the $s$-dimensional Euclidean space $\mathbb{R}^s$.  The map $\nu_{\mc D}$ need not be onto $\mathbb{R}^s.$ We can then define the `distance' function as $\|\nu_\mathcal{D}(\theta)\| = \| \nu_{\mc D}(\theta) \| _1 =\sum_{j=1}^s | \nu_j(\theta)|$.  We use `distance' to make clear that $\|\nu_{\mc D}(\theta)\|$ will not in general satisfy the requirements of a metric on $\mc R$.  

The constraint functions for $\mc D$ cannot be unique; one may always replace the constraint $\nu_j(\theta) = 0$ with the equivalent constraint $k\nu_j(\theta)= 0 $ for some $k \in \mathbb{R}\setminus \{0\}$.  Additionally, while it is desirable for $\|\nu_\mathcal{D}(\theta)\|$ to be increasing as $\theta$ moves away from the restricted region $\mc D$, it is possible to shrink the posterior towards $\mc D$ more strongly in some directions rather than others by modifying, linearly or nonlinearly, one or more constraint functions without changing the solution set.  Thus, one has a large degree of control in choosing $\|\nu_\mathcal{D}(\theta)\|$ so long as a minimal condition is met, namely $\|\nu_\mathcal{D}(\theta)\|$ is zero for $\theta \in \mathcal{D}$ and positive for $\theta\not\in\mathcal{D}$.

To understand the effects of modifying one or more constraints, from a geometrical perspective, one may use the $d$-expansion of $\mc D$ with respect to $\|\nu_\mathcal{D}(\theta)\|$ which is denoted as 
$${\mc D}_{\|\nu_\mathcal{D}(\theta)\|}(d) = \{ \theta \in \mc R: \|\nu_\mathcal{D}(\theta)\| \le d \}.$$
The $d$-expansion of $\mc D$ is the set of $\theta$ values whose image under $\nu_{\mc D}$ is within $d$ of the origin in  $\mathbb{R}^s$. Equivalently, it is the union of all preimages of $\nu_{\mc D}^{(-1)}(x)$ for $x \in\mathbb{R}^s$ such that $\|x\|_1 \le d.$ Modifying the choice of distance will lead to changes in these preimages and therefore changes in the shape of 
${\mc D}_{\|\nu_\mathcal{D}(\theta)\|}(d)$. 

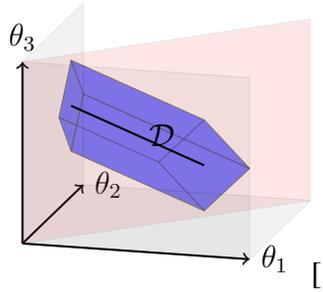
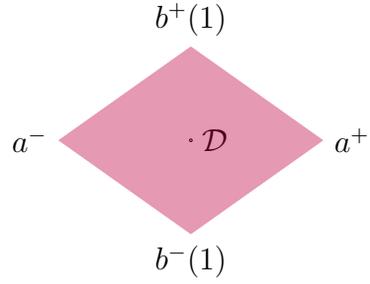
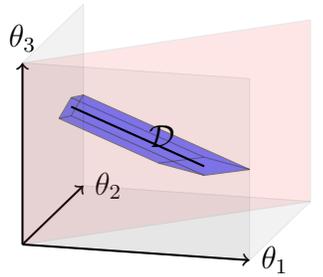
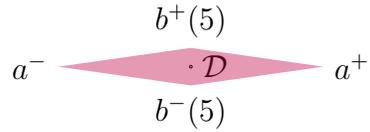
\begin{figure}
\begin{center}
\begin{subfigure}[b]{0.45\textwidth}
\tdplotsetmaincoords{75}{15}
\begin{tikzpicture}[tdplot_main_coords,scale=2.5, line join=bevel]
\def\d{0.25};
\def\k{1};
\draw[thick,->] (0,0,0) -- (1+\d, 0, 0) node[anchor =west]{$\theta_1$};
\draw[thick,->] (0,0,0) -- (0, 1+\d, 0) node[anchor = west]{$\theta_2$};
\draw[thick,->] (0,0,0) -- (0, 0, 1) node[anchor =south]{$\theta_3$};
\draw[fill = red, opacity=0.1] (0,0,0) -- (1+\d,1+\d,0) -- (1+\d,1+\d,1) -- (0,0,1) -- (0,0,0);
\draw[fill = blue, opacity = 0.3]  (1,0,1/2+\d/\k) -- (1+\d,0,1/2) -- (1,0,1/2-\d/\k) --  (1-\d, 0, 1/2)--(1,0,1/2+\d/\k);
\draw[fill = blue, opacity = 0.3] (0,1,1/2+\d/\k) -- (0,1+\d,1/2) -- (0,1,1/2-\d/\k) --  (0, 1-\d, 1/2)--(0,1,1/2+\d/\k) ; 
\draw[fill = blue, opacity = 0.3] (1-\d,0,1/2) -- (0,1-\d,1/2) -- (0,1,1/2+\d/\k) -- (1,0,1/2+\d/\k) -- (1-\d,0,1/2);
\draw[fill = blue, opacity = 0.3] (1-\d,0,1/2) -- (0,1-\d,1/2) -- (0,1,1/2-\d/\k) -- (1,0,1/2-\d/\k) -- (1-\d,0,1/2);
\draw[fill = blue, opacity = 0.3] (1+\d,0,1/2) -- (0,1+\d,1/2) -- (0,1,1/2+\d/\k) -- (1,0,1/2+\d/\k) -- (1+\d,0,1/2);
\draw[fill = blue, opacity = 0.3] (1+\d,0,1/2) -- (0,1+\d,1/2) -- (0,1,1/2-\d/\k) -- (1,0,1/2-\d/\k) -- (1+\d,0,1/2);
\draw[fill = gray, opacity = 0.1] (0,0,0) -- (1+\d,0,0) -- (1+\d,0,1) -- (0,0,1) -- (0,0,0);
\draw[fill = gray, opacity = 0.1] (0,0,0) -- (0,1+\d,0) -- (0,1+\d,1) -- (0,0,1) -- (0,0,0);
\draw[fill = gray, opacity = 0.1] (0,0,0) -- (1+\d,0,0) -- (1+\d,1+\d,0) -- (0,1+\d,0) -- (0,0,0);
\draw[thick] (0,1,1/2) -- (1/2,1/2,1/2)  node[anchor = west]{$\mc D$} -- (1,0,1/2) ;
\end{tikzpicture}[
\caption{$d=1/4,\, k=1$}
\end{subfigure}
\quad
\begin{subfigure}[b]{0.45\textwidth}
\begin{tikzpicture}[scale=2.5, line join=bevel]
\draw [color = white] (0,-1/4) -- (0,1/2);
\def\d{0.5};
\def\k{1};
\node (a) at (-1.41*\d,1/2) [anchor = east]{$a^-$};
\node (b) at (0,1/2+\d/\k) [anchor = south]{$b^+(1)$};
\node (c) at (1.41*\d,1/2) [anchor = west]{$a^+$};
\node (d) at (0,1/2-\d/\k) [anchor = north]{$b^-(1)$};
\draw (0,1/2) circle[radius=0.2pt,fill = black];
\node  at (0,1/2) [anchor = west]{$\mc D$};
\path[fill=purple,opacity = 0.4] (-1.41*\d,1/2) to (0,1/2+\d/\k)  to (1.41*\d,1/2) to(0,1/2-\d/\k)  to  (-1.41*\d,1/2) ;
\end{tikzpicture}
\caption{$d=1/4,\, k=1$}
\end{subfigure}

\begin{subfigure}[b]{0.45\textwidth}
\tdplotsetmaincoords{75}{15}
\begin{tikzpicture}[tdplot_main_coords,scale=2.5, line join=bevel]
\def\d{0.25};
\def\k{5};
\draw[thick,->] (0,0,0) -- (1+\d, 0, 0) node[anchor =west]{$\theta_1$};
\draw[thick,->] (0,0,0) -- (0, 1+\d, 0) node[anchor = west]{$\theta_2$};
\draw[thick,->] (0,0,0) -- (0, 0, 1) node[anchor =south]{$\theta_3$};
\draw[fill = red, opacity=0.1] (0,0,0) -- (1+\d,1+\d,0) -- (1+\d,1+\d,1) -- (0,0,1) -- (0,0,0);
\draw[fill = blue, opacity = 0.3]  (1,0,1/2+\d/\k) -- (1+\d,0,1/2) -- (1,0,1/2-\d/\k) --  (1-\d, 0, 1/2)--(1,0,1/2+\d/\k);
\draw[fill = blue, opacity = 0.3] (0,1,1/2+\d/\k) -- (0,1+\d,1/2) -- (0,1,1/2-\d/\k) --  (0, 1-\d, 1/2)--(0,1,1/2+\d/\k) ; 
\draw[fill = blue, opacity = 0.3] (1-\d,0,1/2) -- (0,1-\d,1/2) -- (0,1,1/2+\d/\k) -- (1,0,1/2+\d/\k) -- (1-\d,0,1/2);
\draw[fill = blue, opacity = 0.3] (1-\d,0,1/2) -- (0,1-\d,1/2) -- (0,1,1/2-\d/\k) -- (1,0,1/2-\d/\k) -- (1-\d,0,1/2);
\draw[fill = blue, opacity = 0.3] (1+\d,0,1/2) -- (0,1+\d,1/2) -- (0,1,1/2+\d/\k) -- (1,0,1/2+\d/\k) -- (1+\d,0,1/2);
\draw[fill = blue, opacity = 0.3] (1+\d,0,1/2) -- (0,1+\d,1/2) -- (0,1,1/2-\d/\k) -- (1,0,1/2-\d/\k) -- (1+\d,0,1/2);
\draw[fill = gray, opacity = 0.1] (0,0,0) -- (1+\d,0,0) -- (1+\d,0,1) -- (0,0,1) -- (0,0,0);
\draw[fill = gray, opacity = 0.1] (0,0,0) -- (0,1+\d,0) -- (0,1+\d,1) -- (0,0,1) -- (0,0,0);
\draw[fill = gray, opacity = 0.1] (0,0,0) -- (1+\d,0,0) -- (1+\d,1+\d,0) -- (0,1+\d,0) -- (0,0,0);
\draw[thick] (0,1,1/2) -- (1/2,1/2,1/2)  node[anchor = west]{$\mc D$} -- (1,0,1/2) ;
\end{tikzpicture}
\caption{$d=1/4,\, k=5$}
\end{subfigure}
\quad
\begin{subfigure}[b]{0.45\textwidth}
\begin{tikzpicture}[scale=2.5, line join=bevel]
\draw [color = white] (0,-1/4) -- (0,1/2);
\def\d{0.5};
\def\k{5};
\node (a) at (-1.41*\d,1/2) [anchor = east]{$a^-$};
\node (b) at (0,1/2+\d/\k) [anchor = south]{$b^+(5)$};
\node (c) at (1.41*\d,1/2) [anchor = west]{$a^+$};
\node (d) at (0,1/2-\d/\k) [anchor = north]{$b^-(5)$};
\draw (0,1/2) circle[radius=0.2pt,fill = black];
\node  at (0,1/2) [anchor = west]{$\mc D$};
\path[fill=purple,opacity = 0.4] (-1.41*\d,1/2) to (0,1/2+\d/\k)  to (1.41*\d,1/2) to(0,1/2-\d/\k)  to  (-1.41*\d,1/2) ;
\end{tikzpicture}
\caption{$d=1/4,\, k=5$}
\end{subfigure}
\end{center}
\caption{The $d$-expansion arising from \eqref{EQ:Constraint_Example} is shown for $d=1/4$ in the positive orthant of $\mathbb{R}^3$. The constraint, $\mc D$, is shown in black.   In the left column, the $d$-expansion is shown in blue.  For reference the plane $\theta_1 - \theta_2 =0$ is shown in red and its intersection with $d$-expansion is shown in purple in the right column.  The increased shrinkage in the $\theta_3$ direction, as $k$ is increased from one to five, is clear.  Here, $a^\pm=\big(1/2\pm d2^{-1/2},1/2\pm d2^{-1/2},1/2\big) $ and $b^\pm(k) =\big(1/2,1/2,1/2\pm d/k\big)$. }
\label{FIG:d_Expansion}
\end{figure}

To expand on this idea further consider the following example. Suppose $\theta \in \mathbb{R}^3=\mc R$ is constrained to the line $\mc D = \{\theta = (\theta_1,\theta_2,\theta_3) : \theta_1 + \theta_2 = 1,\, \theta_3=1/2\}$.  For $k>0$ consider the constraint functions 
\begin{equation*}
\nu_1(\theta) = \theta_1 + \theta_2 - 1, \hspace{5 mm}
\nu_2(\theta) = k(\theta_3 -1/2) 
\end{equation*}
so that 
\begin{equation}
\|\nu_{\mc D}(\theta) \| = | \theta_1 + \theta_2 - 1| + k |\theta_3 -1/2|.
\label{EQ:Constraint_Example} 
\end{equation}  As $k$ is increased, $\|\nu_{\mc D}(\theta)\|$ will shrink $\theta$ towards $\mc D$ more strongly in the $\theta_3$ direction.  Examples of the $d$-expansion reflecting this effect are shown in Fig. \ref{FIG:d_Expansion}.

The distance should then be chosen based on prior belief about how the probability should decrease outside of $\mc D$.  In the absence of prior knowledge supporting one choice over another, one can potentially try several different choices, while assessing sensitivity of the results.  Even though the precise shape of the $d$-expanded regions, and hence the tails of the prior density outside of $\mc D$, can depend on the choice of distance, results are typically essentially indistinguishable practically for different choices.  Differences in the $d$-expanded regions, such as those shown in Figure~\ref{FIG:d_Expansion}, tend to lead to very minimal differences in posterior inferences when $\lambda$ is small in our experience.

\subsection{Constructing the Relaxed Posterior}
\label{SEC:Zero_Measure_Methods}

As $\mathcal{D}$ is a measure
zero subset of $\mathcal{R}$, the sharply constrained posterior density cannot be constructed by truncating the unconstrained posterior on $\mc D$ and renormalizing.  Rather, one must first use techniques from geometric measure theory to define a regular conditional probability for sharp constraints on $\mc D$. This additional step gives rise to a number of technical difficulties, discussed in more detail in Section 3.  Most notably, the use of regular conditional probability restricts the formulation of the distance function $\| \nu_{\mc D}(\theta)\|$ used in constraint relaxation.  

As a guiding example, let us return to the line from the previous section where $\mc D = \{\theta: \theta_1+\theta_2=1,\, \theta_3 = 1/2\}$ and we assume a prior density $\pi_{\mc R}(\theta) = \mathbbm{1}_{(0,1)^3}(\theta)$ which is uniform distribution on the unit cube.  As $\mu_{\mc R}(\mc D)=0$, the sharply constrained posterior \eqref{eq:sharp}
cannot be absolutely continuous with respect to $\mu_{\mc R}$.  To circumvent this issue, one possibility is to replace
 $(\theta_1,\theta_2,\theta_3)$ with $(\theta_1,1-\theta_1,1/2)$, reducing the dimension of the problem. This approach is equivalent to building a regular conditional probability on $\mc D$ which results in a posterior density defined with respect to the normalized 1-Hausdorff measure, or arclength, on $\mc D$. In this reparameterized lower dimensional setting, the constraint is strictly enforced, eliminating any relaxation away from $\mc D$.  
 
Alternatively, one can create a relaxed posterior in the following manner.
Motivated by the original constraints in the specification of $\mc D$, $\theta_1+\theta_2=1$ and $\theta_3 = 1/2$, we set $\nu_{\mc
  D}(\theta) = |\theta_1+\theta_2 - 1| + |\theta_3-1/2|$ so that $\| \nu_{\mc D}(\theta) \| =0$
when $\theta \in \mc D$ and is positive otherwise. 
We then define the relaxed posterior as
\begin{equation*}
  \begin{aligned}
\tilde{\pi}_\lambda(\theta) & \propto \mathcal{L}(\theta;Y)
\pi_\mc R (\theta) \exp\bigg(-\frac{\| \nu_{\mc D} (\theta)\|}{\lambda} \bigg) \\
&=
\mathcal{L}(\theta;Y)  \exp\bigg(- \frac{\| \theta_1 + \theta_2
  -1\| +|\theta_3 - 1/2|}{\lambda}\bigg) \alex{\mathbbm{1}_{(0,1)^3}(\theta_1, \theta_2,\theta_3)}.
  \end{aligned}
\end{equation*}
This density, defined with respect to Lebesgue measure on the plane, makes use of the original constraints, $\theta_1+\theta_1=1$ and $\theta_3=1/2$, to define a `distance' function which in turn allows for relaxation away from the constrained space.

More generally, assume that $\mathcal{D}$ has zero $r$-dimensional Lebesgue measure, corresponding to zero volume within $\mathcal{R}$ and that it may be defined implicitly as the solution set of $\nu_{\mc D}(\theta) = [\nu_1(\theta),\dots,\nu_s(\theta)]^t = 0$.  Under some mild assumptions on the constraint functions, the preimages, $\nu_{\mc D}^{(-1)}(x)$, will be $(r-s)$-dimensional submanifolds of $\mc R$ for $\mu_{\mathbb{R}^s}$-almost every $x$ in the range of $\nu_{\mc D}.$ In particular, $\nu_{\mc D}^{(-1)}(0) = \mc D.$ 
Thus, we expect that $\mc D$ will have  non-zero $(r-s)$-dimensional surface area, corresponding to the 
normalized $(r-s)$-dimensional Hausdorff measure, denoted by $\bar{\mathcal{H}}^{(r-s)}$.  To construct the sharply constrained posterior density, we renormalize the fully constrained density by its integral with respect to the normalized $(r-s)$-dimensional Hausdorff measure yielding a regular conditional probability on the constrained space. 
Using the normalized Hausdorff measure, we take
$$\pi_{\mc D}(\theta \mid Y) = \dfrac{
{\mathcal{L}(\theta;Y)\pi_\mathcal{R}(\theta)}{J^{-1}(\nu_\mathcal{D}(\theta))
\mathbbm{1}_{\mc D}(\theta)
}
} {\int_\mathcal{D}
{\mathcal{L}(\theta;Y)\pi_\mathcal{R}(\theta)}{J^{-1}(\nu_\mathcal{D}(\theta))}
d\bar{\mathcal{H}}^{(r-s)}(\theta)} \propto
{\mathcal{L}(\theta;Y)\pi_\mathcal{R}(\theta)}{J^{-1}(\nu_\mathcal{D}(\theta))
\mathbbm{1}_{\mc D}(\theta)
},
$$
where the Jacobian of $\nu_{\mc D}$, $J(\nu_\mathcal{D}(\theta)) =
[(D\nu_\mathcal{D})^T(D\nu_\mathcal{D})]^{1/2}$, is assumed to be positive and arises from the co-area formula \citep{federer2014geometric}.  The Jacobian, in part, accounts for the differences in dimension between $\mc D$ and $\mc R.$

To relax the constraint we begin with \eqref{eq:sharp} and replace the
indicator function with $\exp(-\|\nu_\mathcal{D} (\theta)\|/\lambda)$, adding
support for  $\|\nu_\mathcal{D}(\theta)\|>0$.  Therefore, the relaxed density is 
\begin{equation}
\label{EQ:relaxedDensityZeroMeasure}
 \tilde{\pi}_\lambda (\theta) \propto \mathcal{L}(\theta;Y)\pi_{\mc R}(\theta) \exp\bigg(-\lambda^{-1} \| \nu_\mathcal{D}(\theta) \|\bigg) =\mathcal{L}(\theta;Y)\pi_{\mc R}(\theta) \exp\bigg(-\lambda^{-1}  \sum_{j=1}^s \| \nu_j(\theta)\|\bigg) .
 \end{equation} Unlike the sharply constrained density however, the relaxed density is supported on $\mc R$ and is defined with respect to $\mu_{\mc R}$.  One important result of this difference is that the Jacobian of $\nu_{\mc D}$ does not appear in \eqref{EQ:relaxedDensityZeroMeasure}. 
As a result, the sharply constrained posterior density is not a pointwise
limit of the relaxed density in general.  For a more complex example, see the Supplementary Materials for constraint relaxed modeling on the torus in $\mathbb{R}^3$.

\section{Theory}

%
%
%
%
\label{SEC:Zero_measure_theory}

We begin with a review of some important concepts of
geometric measure theory. In addition to supporting the analysis, we are reviewing these topics to offer insight into the behavior of the relaxed posterior.  
 \begin{definition} Let
$A\subset \bb R^r$. Fix $d \le r$. \alex{The $d$-dimensional Hausdorff measure of $A$ is} $$\mc H^{d}(A)=
\underset{\delta\rightarrow 0}\lim \inf \bigg\{ \sum
\left[{\text{diam}(S_i)}\right]^d: {A\subseteq \bigcup S_i,
\text{diam}(S_i)\le \delta}, \text{diam}(S_i)=\sup_{x,y\in
S}\|x-y\|\bigg\}.$$  \end{definition} 
We denote the normalized $d$-dimensional Hausdorff
measure as $$\bar{\mc H}^{d}(A) =\frac{\Gamma(\frac{1}{2})^{d}}{2^d
\Gamma(\frac{d}{2}+1)} \mc H^{d}(A).$$ When $d=r$, Lebesgue and normalized
Hausdorff measures coincide,  $\mu_{\mathbb{R}^r}(A)= \bar{\mc H}^{d}(A)$
\citep{evans2015measure}.  Additionally, for a subset $\mathcal{D}$, there
exists a unique, critical value $d$ such that
$\bar{\mathcal{H}}^s(\mathcal{D}) = 0$ for $s>d$ and $\infty$ for $s<d$.
The critical value, $d$, is referred to as the Hausdorff dimension of $\mathcal{D}$, which agrees with the usual notion of dimension when $\mc D$ is a piecewise smooth manifold. In fact, when $\mathcal{D}$ is a
compact, $d$-dimensional submanifold of $\mathbb{R}^m$, it will have
Hausdorff dimension $d$ so that $\bar{\mathcal{H}}^d(\mathcal{D})$ is the
$d$-dimensional surface area of $A.$ As discussed in Section 2, we are focusing on the case where $\mc D$ is an $(r-s)$-dimensional submanifold of $\mc R$. As such, it is natural to define the sharply constrained posterior with respect to $\bar{\mc H}^{r-s}$, which is referred to as a regular conditional probability (\cite{diaconis2013manifold}).

Defining the regular conditional probability on the measure zero constrained space $\mathcal{D}$
and the subsequent analysis requires the co-area formula.
\begin{theorem}{Co-area formula \citep{diaconis2013manifold,
federer2014geometric}} Suppose $v:\mathbb{R}^r\to\mathbb{R}^s$,
with $s<r$, is Lipschitz and that
$g\in\mathbb{L}^1(\mathbb{R}^r,\mu_{\mathbb{R}^r}).$ Assume
$J[v(\theta)]>0$, then \begin{equation} \int_{\mathbb{R}^r}
g(\theta)J[v(\theta)]d\mu_{\mathbb{R}^r}( \theta)=
\int_{\mathbb{R}^s} \bigg( \int_{v^{-1}(y)}g(\theta)
d\bar{\mathcal{H}}^{r-s}(\theta)\bigg)d\mu_{\mathbb{R}^s}(y).
\end{equation} \end{theorem} Recall, we previously assumed that $\mathcal{D}$ can be defined implicitly as the
solution set to a system of $s$ equations,  $\{\nu_j(\theta)=0\}_{j=1}^s$, and we defined the map $\nu_{\mc D}(\theta) = [\nu_1(\theta),\dots,\nu_s(\theta)]$ from our parameter space, $\mc R$, to the Euclidean space, $ \mathbb{R}^s.$ These constraint functions must adhere to some additional restrictions: \alex{(\emph{a}) $\nu_j:\mathcal{R}\to\mathbb{R}$ is
Lipschitz continuous, (\emph{b}) $\nu_j(\theta)=0$ only for
$\theta\in\mathcal{D}$, (\emph{c}) for $j=1,\dots, s$, the
pre-image
$\nu_j^{(-1)}(x)$ is a co-dimension 1 sub-manifold of $\mathcal{R}$
for $\mu_\mathbb{R}$-almost every $x$ in the range of $\nu_j$, and (\emph{d})
$\nu_j^{(-1)}(0)$ and $\nu_k^{(-1)}(0)$ intersect transversally for
$1\le j<k\le s.$ }

Property (\emph{a}) guarantees that $\nu_{\mc D}$ is itself Lipschitz so the co-area formula applies.  The remaining
properties (\emph{b})-(\emph{d}) are constructed so that when $x\in\mathbb{R}^s$ is near zero, the preimage $\nu_{\mc D}^{(-1)}(x)$ is also an $(r-s)$-dimensional submanifold corresponding to a perturbation of the constrained space $\mc D$. In the remainder of this section, we assume that $\nu_{\mc D}^{(-1)}(x)$ is an $(r-s)$-dimensional submanifold of $\mu_{\mathbb{R}^s}$ almost every $x$ in the range of $\nu_\mc D$.  While this is a very strong assumption, to attain relaxation near $\mc D$, the transversality condition (d) assures this for $x$ near 0.

\alex{While (\emph{a}) - (\emph{d}) may seem restrictive, many measure zero constraints can be defined implicitly to satisfy them}.  In Table \ref{TABLE:Equality_constraints_examples}, we offer a few examples. 
An initial set of constraint functions can typically be modified to satisfy the Lipschitz condition by truncating the 
original parameter space $\mc R$ or by composing the constraints with bounded functions. The former  choice was used for the Unit sphere and Stiefel manifold constraints in the table.

\begin{table}[h!]
\begin{center} 
\begin{tabular}{| c | p{4cm} | c | c | p{4cm} |}
\hline $\mathcal{R}$                           &
$\mathcal{D}$                                  &
$\dim(\mc R)$                                  &
$\dim( \mc D)$                                 & Constraint functions                                  \\ \hline $[0,1]^r$ &
Probability simplex, $\Delta^{r-1}$            & $r$                            & $r-1$ & $\nu_1(\theta)
= \sum(\theta) -1$                                                                                     \\ \hline $\mathbb{R}^r$ & Line,
span$\{\vec{u}\}$ \newline $\vec{u}\ne\vec{0}$ & $r$                            & $1$
&
$\nu_j(\vec{\theta}) = \vec{\theta}\,^T\vec{b}_j$
$\{\vec{b}_1,\dots,\vec{b}_{r-1}\}$ a basis for
span$\{\vec{u}\}^\perp$                                                                                \\ \hline $[-1,1]^r$ & Unit
sphere, $\mathbb{S}^{r-1}$                     & $r$                            & $r-1$ & $\nu_1(\theta) =
\|\theta\|^2 -1$                                                                                     \\ \hline $[-1,1]^{n\times
k}$   & Stiefel manifold, $\mc V(n,k)$ & $nk$  & $nk -
\binom{k+1}{2}$                             & $\nu_{i,j}(\theta) = 
\vec{\theta}_i'\vec{\theta}_j- \delta_{i,j}$ \newline
$1\le i \le j \le k$ and $\delta_{i,j} = \mathbbm{1}_{i=j}$
\\ \hline\end{tabular} \end{center} \caption{Table of
constraints for some commonly used constrained spaces.}
\label{TABLE:Equality_constraints_examples} \end{table}


Given this construction of the constrained space, we can now specify the regular conditional probability of $\theta$, given $\theta \in
\mathcal{D}.$

\begin{theorem}\citep{diaconis2013manifold}
\label{THM:RCP_construction} Assume that $J(\nu_{\mc D}(\theta)) > 0$ and
that for each $z\in\mathbb{R}^s$ there is a finite non-negative
$p_z$ such that,  $$m^{p_z}(z) = \int_{\nu_{\mc D}^{-1}(z)}
\frac{\mathcal{L}(\theta; Y) \pi_\mathcal{R}(\theta)}
{J(\nu_{\mc D}(\theta))}
d\bar{\mathcal{H}}^{p_z}(\theta)
\in (0,\infty).$$
Then, for any Borel subset $F$ of
$\mathcal{R}$, it follows that $$P(\theta \in F \mid v(\theta) = z) = \begin{cases}
\frac{1}{m^{p_z}(z)} \int_{F} \frac{\mathcal{L}(\theta; Y)
\pi_\mathcal{R}(\theta)
\mathbbm{1}_{\nu_{\mc D}(\theta)=z}}{J(\nu_{\mc D}(\theta))}
d\bar{\mathcal{H}}^{p_z}(\theta)
& m^p(z)\in (0,\infty) \\ \delta (F) & m^p(z) \in \{0,\infty\}
\end{cases}$$
is a valid regular conditional probability for
$\theta\in\mathcal{D}.$ Here, $\delta (F)=1$ if $0\in F$ and $0$ otherwise.
\end{theorem}

By construction, $\{\theta:\nu_{\mc D}(\theta)=z\}$ is an $(r-s)$ dimensional
submanifold of $\mathcal{R}$ for $\mu_{\mathbb{R}^s}$ almost every $z$ in
the range of $\nu_{\mc D}$. It follows that one should take $p_z=r-s$. Most importantly, setting $z=0$ allows us to define

\begin{equation} \label{EQ:Constrained_rcp}
\pi_\mathcal{D}(\theta\mid\theta\in\mathcal{D},Y) = \frac{1}{m^{r-s}({0})}
\frac{\mathcal{L}(\theta; Y) \pi_\mathcal{R}(\theta)
\mathbbm{1}_{\mc D}(\theta)}{J(\nu_{\mc D}(\theta))} \end{equation} as the
constrained posterior density as originally stated in Section 2.2.  

To understand the effects of constraint relaxation, consider a Borel subset, $\mathcal{F}$, of $\mc R.$ Under the sharply constrained posterior,
\begin{equation}
\begin{aligned}
P(\theta\in\mc F\mid Y) &=\int_{\mc F} \pi_{\mc D}(\theta\mid Y) d\bar{\mc H}^{r-s}(\theta) =\frac{\int_{\mc F}  \mathcal{L}(\theta;Y)\pi_\mathcal{R}(\theta)J^{-1}(\nu_\mathcal{D}(\theta)) \mathbbm{1}_{\mc D}(\theta) d\bar{\mc H}^{r-s}(\theta)}{\int_{\mc D}  \mathcal{L}(\theta;Y)\pi_\mathcal{R}(\theta)J^{-1}(\nu_\mathcal{D}(\theta)) \mathbbm{1}_{\mc D}(\theta) d\bar{\mc H}^{r-s}(\theta)}\\
&=\frac{\int_{\mc F\cap \mc D}  \mathcal{L}(\theta;Y)\pi_\mathcal{R}(\theta)J^{-1}(\nu_\mathcal{D}(\theta)) d\bar{\mc H}^{r-s}(\theta)}{\int_{\mc D}  \mathcal{L}(\theta;Y)\pi_\mathcal{R}(\theta)J^{-1}(\nu_\mathcal{D}(\theta)) d\bar{\mc H}^{r-s}(\theta)}.
\end{aligned}
\end{equation}
Alternatively, under the relaxed posterior,
\begin{equation}
\begin{aligned}
P(\theta \in \mc F \mid Y) &= \int_{\mc F} \tilde{\pi}_\lambda(\theta) d\mu_{\mc R}(\theta) = \frac{\int_{\mc F} \mathcal{L}(\theta;Y)\pi_{\mc R}(\theta) \exp\bigg(-\lambda^{-1}  \sum_{j=1}^s \| \nu_j(\theta)\|\bigg) d\mu_{\mc R}(\theta) }{\int_{\mc R} \mathcal{L}(\theta;Y)\pi_{\mc R}(\theta) \exp\bigg(-\lambda^{-1}   \sum_{j=1}^s \| \nu_j(\theta)\|\bigg) d\mu_{\mc R}(\theta)}
\end{aligned}
\label{EQ:measure_zero_relax}
\end{equation}

Making use of the behavior of the preimages of $\nu_{\mc D}$, we can reexpress \eqref{EQ:measure_zero_relax} through the co-area formula as 
\begin{equation}
\begin{aligned}
  P(\theta \in \mc F \mid Y)
= \frac{\int_{\mathbb{R}^s} \bigg[\int_{\mc F \cap \,\nu_{\mc D}^{(-1)}(x)   } \mathcal{L}(\theta;Y)\pi_{\mc R}(\theta) J^{-1}(\nu_\mathcal{D}(\theta)) d\bar{\mc H}^{r-s}(\theta)  \bigg] \exp\bigg(-\lambda^{-1}  \|x\|_1 \bigg) d\mu_{\mathbb{R}^s }(x) }{\int_{\mathbb{R}^s}\bigg[\int_{\mc R \cap \,\nu_{\mc D}^{(-1)}(x)   } \mathcal{L}(\theta;Y)\pi_{\mc R}(\theta) J^{-1}(\nu_\mathcal{D}(\theta)) d\bar{\mc H}^{r-s}(\theta)  \bigg] \exp\bigg(-\lambda^{-1}  \|x\|_1\bigg)d\mu_{\mathbb{R}^s }(x) } 
\end{aligned}
\end{equation}


The posterior expectation of $g(\theta)$ under the sharp constraint $\theta \in
\mathcal{D}$ is
$$E[g(\theta) \mid \theta\in\mathcal{D}] = E[g(\theta) \mid
\nu_{\mc D}(\theta) =0\,] = \int_\mathcal{R} g(\theta) \pi_\mathcal{D}(\theta)
d\bar{\mathcal{H}}^{r-s}(\theta).$$
Using the definition of $\tilde{\pi}_\lambda$ from Section 2.2, the expected value of
$g(\theta)$ with respect to the relaxed density, denoted
$E_{\tilde{\Pi}}[g(\theta)] $, is

\begin{equation}
E_{\tilde{\Pi}}[g(\theta)] = \frac{1}{m_\lambda}
\int_\mathcal{R} g(\theta) 
\mathcal{L}(\theta;Y)\pi_\mathcal{R}(\theta)\exp\bigg(-\lambda^{-1}\|\nu_{\mc D} (\theta)\|_1\bigg)
d\mu_\mathcal{R}(\theta) 
\label{EQ:Expectation_zero_measure_relaxed}
\end{equation}
where $m_\lambda =
\int_\mathcal{R}  
\mathcal{L}(\theta;Y)\pi_\mathcal{R}(\theta)\exp(-{\lambda^{-1}}\|\nu_{\mc D} (\theta)\|_1)
d\mu_\mathcal{R}(\theta).$
We may now state the primary result regarding
the use of $E_{\tilde{\Pi}}[g(\theta)]$ to estimate $E[g(\theta)\mid\theta\in\mathcal{D}]$.

\begin{theorem} \label{THM:Relaxed_Expectation_Convergence_Measure_Zero}
Let $m:\mathbb{R}^s\to \mathbb{R}$ and $G:\mathbb{R}^s\to
\mathbb{R}$ be defined as follows \begin{align*} m(x) & =
\int_{\nu_{\mc D}^{-1}(x)} \frac{
\mathcal{L}(\theta;Y)\pi_\mathcal{R}(\theta)}{J(\nu_{\mc D}(\theta))}
d\bar{\mathcal{H}}^{r-s}(\theta) \\ G(x) &= \int_{\nu_{\mc D}^{-1}(x)}
g(\theta)\frac{
\mathcal{L}(\theta;Y)\pi_\mathcal{R}(\theta)}{J(\nu_{\mc D}(\theta))} d\bar{\mathcal{H}}^{r-s}(\theta).
\end{align*} Suppose that both $m$ and $G$ are continuous on an open
interval containing the origin and that \\
$g\in\mathbb{L}^1(\mathcal{R},\mathcal{L}(\theta;Y)\pi_{\mc R}(\theta) d\mu_\mathcal{R})$.
Then, $\big|E_{\tilde{\Pi}}[g(\theta)] - E[g(\theta) \mid\theta \in \mathcal{D}]\big| \to 0
\text{ as } \lambda\to 0^+.$ 
\alex{Additionally, if $m$ and $G$ are differentiable at $0$, then
$$\big|E_{\tilde{\Pi}}[g] - E[g(\theta)\mid\theta \in \mathcal{D}] \big| =
O\bigg(\frac{\lambda}{|\log \lambda|^s}\bigg)$$ as $\lambda \to
0^+.$ }
\end{theorem}

The continuity  and differentiability assumptions of Theorem \ref{THM:Relaxed_Expectation_Convergence_Measure_Zero} have some important consequences.  Recall, the pairwise transversal intersection requirement, (\emph{d}), assures that $\nu_{\mc D}^{(-1)}(x)$ behaves like a small perturbation of $\mc D$ when $x$ is near zero. Therefore, if the unconstrained posterior, $\mathcal{L}(\theta;Y)\pi_{\mc R}(\theta)$, the Jacobian, $J(\nu_{\mc D}(\theta))$, and $g$ are all continuous on an open neighborhood containing $\mc D$, the continuity assumptions of Theorem \ref{THM:Relaxed_Expectation_Convergence_Measure_Zero} will follow.  A proof of Theorem \ref{THM:Relaxed_Expectation_Convergence_Measure_Zero} is provided in the appendix. 

%

\section{Posterior Computation}

The constraint relaxed posterior density is supported in $\mc R$ and can be directly
sampled via off--the--shelf tools, such as slice
sampling, adaptive Metropolis-Hastings, and Hamiltonian Monte Carlo.
In this section, we focus on  Hamiltonian Monte Carlo as a general algorithm that tends to have good performance in a variety of settings.

In order to sample $\theta$,  Hamiltonian Monte Carlo introduces an auxiliary momentum variable $p
\sim \No(0, \mass)$. The covariance matrix $\mass$ is referred to as a
\textit{mass matrix} and is typically chosen to be the identity or adapted
to approximate the inverse covariance of $\theta$.  Hamiltonian Monte Carlo then samples from the
joint target density $\pi(\theta, p) = \pi(\theta) \pi(p) \propto \exp \{- H(\theta, p)\}$
where, in the case of the posterior under relaxation,
$H(\theta,p) = U(\theta) + K(p)$, with $U(\theta) = -\log\pi(\theta)$, $K(p) = p'\mass^{-1} p/2$, 
and $\pi(\theta)$ the unnormalized density in  \eqref{EQ:relaxedDensityZeroMeasure}.

From the current state $(\theta^{(0)},p^{(0)})$, Hamiltonian Monte Carlo generates a Metropolis-Hastings proposal by simulating Hamiltonian dynamics defined by a differential equation:
\begin{equation} \begin{aligned} \label{hamiltonian} \frac{\partial \theta
^{(t)}}{\partial t} & =\frac{\partial H(\theta, p)}{\partial p} =
\mass^{-1}p,                                                      \\ \frac{\partial p^{(t)}}{\partial t}&
=-\frac{\partial H(\theta, p)}{\partial \theta} = -\frac{\partial
U(\theta)}{\partial \theta}.\end{aligned} \end{equation}

The exact solution to \eqref{hamiltonian} is typically intractable but a
valid Metropolis proposal can be generated by numerically approximating
\eqref{hamiltonian} with a reversible and volume-preserving  integrator
\citep{neal2011mcmc}. The standard choice is the \textit{leapfrog}
integrator which approximates the evolution $(\theta^{(t)},p^{(t)}) \to (\theta^{(t +
\dt)},p^{(t + \dt)})$ through the following update equations:

\begin{equation} \begin{aligned} \label{leap-frog}
p \leftarrow p -
\frac{\dt}{2} \frac{\partial U}{\partial  \theta },\quad \theta \leftarrow  \theta
+ \dt \mass^{-1}p,\quad p \leftarrow p -  \frac{\dt}{2}
\frac{\partial U}{\partial  \theta }\end{aligned} \end{equation} Taking
$L$ leapfrog steps from the current state $(\theta^{(0)},p^{(0)})$
generates a proposal $(\theta^{*},p^{*}) \approx (\theta^{(L \dt)},p^{(L
\dt)})$, which is accepted with the probability $$1\wedge \exp
\left( - H(\theta^{*},p^{*}) + H(\theta^{(0)},p^{(0)}))\right)$$
The computing efficiency of  Hamiltonian Monte Carlo
  under different degrees of relaxation is discussed in the Supplementary Materials.

\section{Sphere $t$ Distribution}

\begin{figure}[H]
\begin{subfigure}[b]{0.5\textwidth}
\centering
\includegraphics[width=0.7\textwidth]{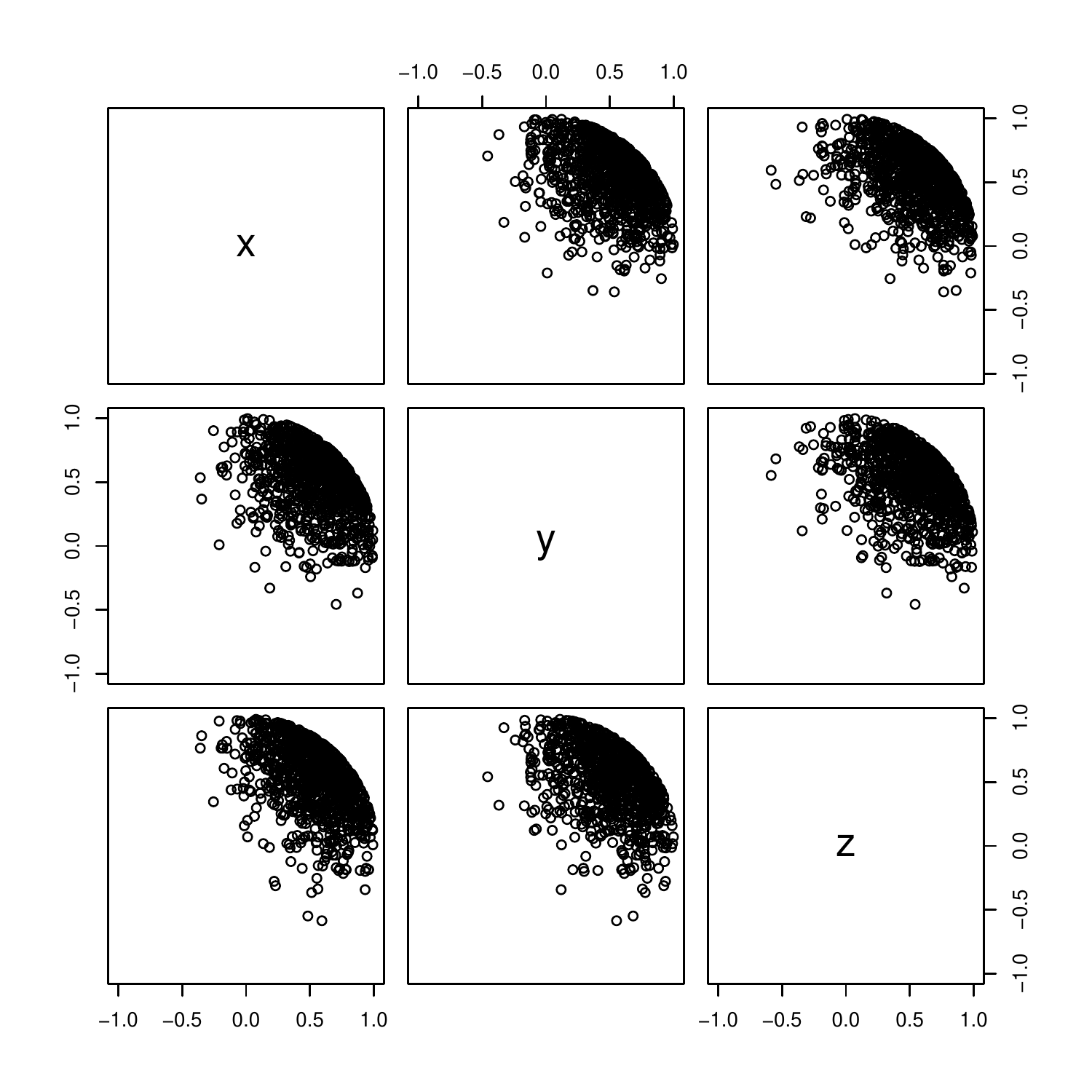}
\caption{von Mises--Fisher distribution.}
\end{subfigure}
\begin{subfigure}[b]{0.5\textwidth}
\centering
\includegraphics[width=0.7\textwidth]{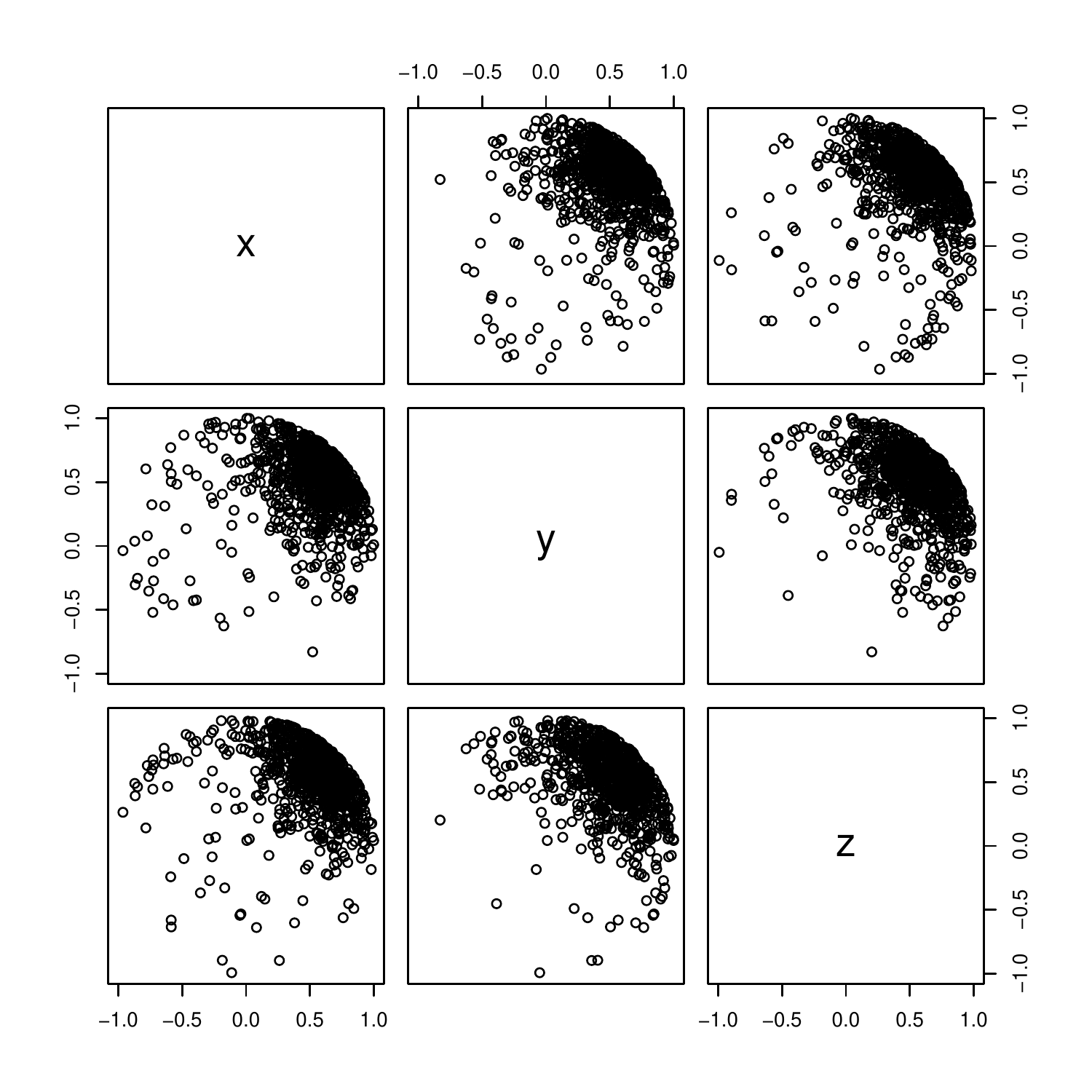}
\caption{Spherical Constraint Relaxed $t$-dist. with $m=3$.}
\end{subfigure}
\caption{Sectional view of random samples from constrained distributions on a
unit sphere inside $\bb R^3$. The distributions are derived through conditioning
on $\theta'\theta=1$ based on unconstrained densities of (a) $\No( F,
\diag\{0.1\})$, (b) $t_3(F,\diag\{0.1\} )$, where
$F=[1/\sqrt{3},1/\sqrt{3},1/\sqrt{3}]'$. The samples are generated via Constraint Relaxed, Hamiltonian Monte Carlo
with $\lambda=10^{-3}$. }
\label{sphere_examples}
\end{figure}

The von
Mises--Fisher distribution \citep{khatri1977mises} is the result of
constraining a multivariate Gaussian $\theta \sim \No(F,I\sigma^2)$ with
$F\in \mc D$ and $v(\theta)= \theta'\theta-1$:
$$
\pi_{\mc D}(\theta) \propto
\exp\bigg(-\frac{\|F-\theta\|^2}{2\sigma^2}\bigg)
\mathbbm{1}_{\theta'\theta=1} 
\propto
\exp\bigg(\frac{F'}{\sigma^2}\theta\bigg)
\mathbbm{1}_{\theta'\theta=1}.$$
Constraint relaxation allows us to easily consider different `parent' unconstrained distributions instead of just the Gaussian. Using the constraint $v(\theta)=\theta'\theta-1$ to form a `distance', we propose a spherical constraint relaxed $t$-density: $$
\tilde\pi_{\lambda}(\theta) \propto \bigg(1+\frac{\|F-\theta\|^2}{m\sigma^2}\bigg)^{-\frac{(m+p)}{2}}\exp\bigg(-\frac{\|\theta'\theta-1\|}{\lambda}\bigg),
$$ with the parent $t$-density having $m$ degrees of freedom, mean $F\in \mc D$ and variance $I\sigma^2$. Figure~\ref{sphere_examples} shows that the proposed distribution induces heavier tail behavior than the von Mises-Fisher.

\section{Sparse Latent Factor Modeling of Brain Networks}

We apply constraint relaxation to analyze brain networks from the KKI-42 dataset 
\citep{landman2011multi}, which consists of two scans for $n=21$ healthy subjects without any
history of neurological disease. For each subject, we take the first scan as the input data, and reserve the second scan for model validation. Data consist of an $R\times R$ symmetric adjacency matrix $A_i$, for $i=1,\ldots,n$, with $R=68$ the number of brain regions and $A_{ikl} \in \{0,1\}$ a 0-1 indicator of 
a connection between regions $k$ and $l$ for individual $i$.

Our focus is on characterizing variation among individuals in their brain networks via a latent factor model, with each factor impacting a subset of the brain regions.  We assume the elements of $A_i$ are conditionally independent given latent factors $v_i = (v_{i1},\ldots,v_{id})'$, with 
\begin{eqnarray}
A_{ikl}  \sim  \mbox{Bern}( \pi_{ikl} ),\quad \log\bigg( \frac{ \pi_{ikl} }{ 1-\pi_{ikl} }\bigg) 
= \mu_{kl} + \psi_{ikl}, \quad \psi_{ikl}  =  \sum_{s=1}^d v_{is} u_{ks}u_{ls}, \label{eq:brainfactor}
\end{eqnarray}
where $\mu_{kl}$ characterizes the overall log-odds of an connection in the $(k,l)$ pair of brain regions, 
and $\psi_{ikl}$ is a subject-specific deviation.  The latent factor $v_{is}$ measures how much individual $i$ expresses brain subnetwork $s$, while $\{ u_{k1},\ldots,u_{kd} \}$ are scores measuring impact of brain region $k$ on the different subnetworks.  We let $\mu_{kl} \sim \No(0,\sigma^2_\mu)$, with 
$\sigma^2_\mu \sim \text{IG}(2,1)$, as a shrinkage prior for the intercept paramters.  We similarly let 
$v_{is} \sim \No(0,\sigma^2_s)$, with $\sigma^2_s \sim \text{IG}(2,1)$, to characterize the population distribution of the $s$th latent factor.

In order for the model to be identifiable, which is important for interpretability, the matrix $U = \{ u_{ks} \}$ needs to be restricted.  A natural constraint is to assume $U \in \mc V(n,d)=\{U: U'U=I_d\}$, corresponding to the Stiefel manifold, to remove rotation and scaling ambiguity \citep{hoff2016equivariant}.  However, there are limited distributional options available on the Stiefel, and it is not clear how to impose sparsity in $U$, so that not all brain regions relate to all latent factors.  To solve this problem, we propose to use a Stiefel constraint relaxed Dirichlet-Laplace shrinkage prior for $U$.  The Dirichlet-Laplace prior was proposed recently as a computationally convenient and theoretically supported prior for incorporating approximate sparsity \citep{bhattacharya2015dirichlet}.  By multipling the Dirichlet-Laplace prior by $\exp(- \lambda^{-1} \|U'U-I\|)$, with $\lambda=10^{-3}$, we obtain a prior that generates realizations $U$ that are very close to orthonormal and sparse.


We compare the resulting model with (i) choosing independent $\No(0,1)$ priors for $u_{ks}$ without constraints; and (ii) choosing such priors with constraint relaxation but no Dirichlet-Laplace shrinkage.  For each model, we run Hamiltonian Monte Carlo for 
$10,000$ iterations and discard the first $5,000$ iterations as burn-in.  For each iteration, we run $300$ leap-frog steps.  We fixed $d=20$ in each case as an upper bound on the number of factors; the shrinkage prior approach can effectively delete factors that are unnecessary for characterizing the data.

\begin{figure}[H] \centering
\includegraphics[width=1\textwidth]{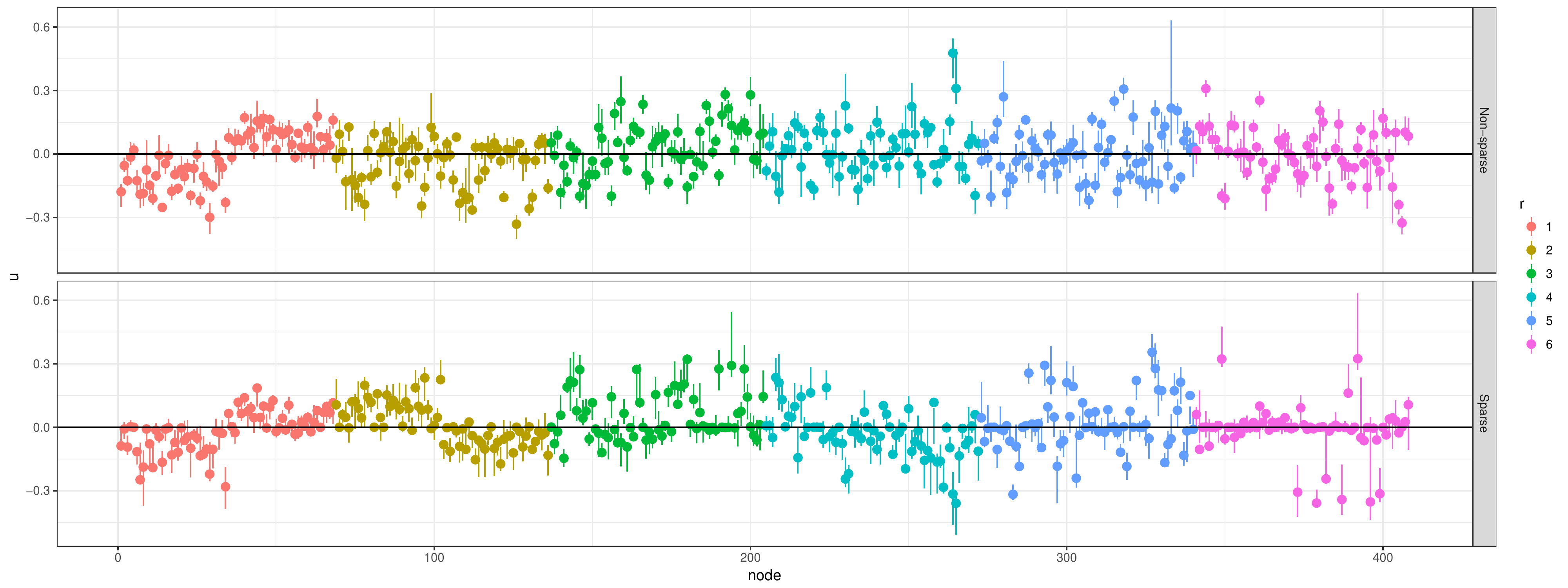}
\caption{Posterior mean and $95\%$ credible intervals of scores
$U_s = \{u_{1s},\ldots,u_{Rs}\}$ for the $R=68$ brain regions, 
and components $s=1,\ldots,6$ (ordered to be decreasing in
$\sigma_s^2$). \label{network_model_basis}}
\end{figure}

As anticipated, the models that did not constrain $U$ failed to converge; one could obtain convergence for identifiable functionals of the parameters, but not directly for the components in the latent factor expansion.  The $\No(0,1)$ and Dirichlet-Laplace shrinkage constraint relaxed models both had good apparent convergence and mixing rates.  Figure~\ref{network_model_basis}(b) plots the top $6$ brain region factors $U_s = \{ u_{1s},\ldots,u_{Rs} \}$ under the normal and shrinkage priors.  The shrinkage prior leads to increasing numbers of brain regions with scores close to zero as the factor index increases.   In addition, shrinkage improves interpretability of the factors; for example, the second factor has positive scores for brain regions in one hemisphere and negative scores for brain regions in the other.

\begin{table}[H] \begin{center} 

\begin{tabular}{ | l || c | c | c|}
\hline
Model & Fitted AUC & Prediction AUC & ESS/1000 iterations \\
\hline
(i) with shrinkage \& near-orthonormality & 97.9\% & 96.2 \% & 193.72 \\
\hline
(ii) with near-orthonormality only & 97.1\% & 96.2 \% & 188.10 \\
\hline
(iii) completely constrainted & 96.9\% & 93.6 \% & 8.15 \\
\hline
\end{tabular}
\end{center} \caption{Benchmark of 3 models for 21 brain networks. Models with
  near-orthonormality show better performance in fitted and prediction AUC,
  in comparing the estimated probability and the network connectivity.
  The unconstrained model has low effective sample size (ESS).
\label{network_model}} \end{table}

We further validate the models by assessing the area under the receiver
operating characteristic curve (AUC). We compute the posterior mean of the 
estimated connectivity probability $\pi_{ikl}$ for each individual and pair
of regions $(k,l)$.  Thresholding these probabilities produces an adjacency
matrix, which we compare with the held-out second scan for each individual.
Table~\ref{network_model} lists the benchmark results. The two models under constraint relaxation
show much better performance, especially in prediction. Shrinkage does not 
improve prediction in this case over imposing orthogonality, but nonetheless is
useful for interpretability as discussed above.

\section{Supplementary Materials}

Supplementary materials include details on positive measure constraints, additional examples, and additional related to the computational efficiency of Hamiltonian Monte Carlo simulation using Constraint Relaxation.



\spacingset{1}
\bibliography{reference} 
\bibliographystyle{chicago}

\appendix
\section{Proof of Main Theorem}
\begin{proof}[of Theorem  \ref{THM:Relaxed_Expectation_Convergence_Measure_Zero}]
For brevity, we use the notation $f(\theta) = \mc L(\theta;Y)\pi_{\mc R}(\theta)$ and $df(\theta) =\mc L(\theta;Y)\pi_{\mc R}(\theta)d\mu_{\mc R}(\theta).$ Recall that we have two densities. The first is the fully constrained density for $\theta\in\mathcal{D}$,
\begin{equation*}
\pi_\mathcal{D}(\theta) = \frac{1}{m_0} \frac{\mathcal{L}(\theta;Y)\pi_\mathcal{R}(\theta)}{J(\nu_{\mc D}(\theta))}\mathbbm{1}_\mathcal{D}(\theta) = \frac{1}{m_0} \frac{f(\theta)}{J(\nu_{\mc D}(\theta))} \mathbbm{1}_{\mc D}(\theta)
\end{equation*}
where the normalizing constant $m_0=\int_\mathcal{R} f(\theta)J^{-1}(\nu_{\mc D}(\theta))\mathbbm{1}_\mathcal{D}(\theta)d\bar{\mathcal{H}}^{r-s}(\theta)$ is calculated with respect to the normalized Hausdorff measure.
Secondly, we have the relaxed distribution
\begin{equation*}\tilde{\pi}_\mathcal{D}(\theta) = \frac{1}{m_\lambda} \mathcal{L}(\theta;Y)\pi_\mathcal{R}(\theta)\exp\bigg(-\frac{\|\nu_{\mc D}(\theta)\|}{\lambda}\bigg)=\frac{1}{m_\lambda} f(\theta)\exp\bigg(-\frac{\|\nu_{\mc D}(\theta)\|}{\lambda}\bigg)\end{equation*}
where the normalizing constant $m_\lambda = \int_\mathcal{R}f(\theta)\exp\big(-\lambda^{-1}\|\nu_{\mc D}(\theta)\|\big) d\mu_\mathcal{R}(\theta)$ is calculated with respect to Lebesgue measure on $\mathcal{R}$.


For a given function, $g:\mathcal{R}\to\mathbb{R}$, we can define the expectation of $g(\theta)$ under the sharp and relaxed posteriors, denoted by $E$ and $E_{\tilde{\Pi}}$ respectively, as
\begin{align*}
&E[g(\theta)|\theta\in\mathcal{D}] = \int_\mathcal{R} \frac{g(\theta)}{m_0} \frac{f(\theta)}{J(\nu_{\mc D}(\theta))}\mathbbm{1}_\mathcal{D}(\theta)d\bar{\mathcal{H}}^{r-s}(\theta)  \\
&E_{\tilde{\Pi}}[g(\theta)]  
=\int_{\mathbb{R}^s} \frac{\exp\big(-\frac{||x||_1}{\lambda}\big)}{m_\lambda}  \int_{\nu_{\mc D}^{-1}(x)} g(\theta) \frac{f(\theta)}{J(\nu_{\mc D}(\theta))} d\bar{\mathcal{H}}^{r-s}(\theta) d\mu_{\mathbb{R}^s}(x). 
\end{align*}
The second equality follows from the co-area formula applied to (14). 
By construction, $m(x)=m^{r-s}(x) = \int_{\nu_{\mc D}^{-1}(x)}f(\theta)J^{-1}(\nu_{\mc D}(\theta)) d\bar{\mathcal{H}}^{r-s}(\theta)> 0$ for $\mu_{\mathbb{R}^s}$-almost every $x\in \text{Range}(\nu_{\mc D})$. Notably, $m_0=m(0)>0$. By Theorem 3,
\begin{equation}
E[g(\theta) | \nu_{\mc D}(\theta) = x] = \frac{1}{m(x)} \int_{\nu_{\mc D}^{-1}(x)} g(\theta)\frac{f(\theta)}{J(\nu_{\mc D}(\theta))} d\bar{\mathcal{H}}^{r-s}(\theta) = \frac{G(x)}{m(x)}.
\end{equation}
As such, we may express $E_{\tilde{\Pi}}[g(\theta)]$ as 
\begin{equation}
E_{\tilde{\Pi}}[g(\theta)] = \int_{\mathbb{R}^s} \frac{m(x)}{m_\lambda}\exp\bigg(-\frac{||x||_1}{\lambda}\bigg) E\big[g(\theta)|\nu_{\mc D}(\theta)=x\big] d\mu_{\mathbb{R}^s}(x) .
\end{equation}

%
%
%
%
%
Let us first consider the small $\lambda$ behavior of $m_\lambda.$ We begin by re-expressing $m_\lambda$ in terms of $m(x)$ through the co-area formula.
\begin{align*}
m_\lambda &= \int_{\mathbb{R}^s} \exp\bigg(-\frac{||x||_1}{\lambda}\bigg) \int_{\nu_{\mc D}^{-1}(x)} \frac{f(\theta) }{J(\nu_{\mc D}(\theta))} d\bar{\mathcal{H}}^{r-s}(\theta) d\mu_{\mathbb{R}^s} (x) =\int_{\mathbb{R}^s}m(x) \exp\bigg(-\frac{||x||_1}{\lambda}\bigg)d\mu_{\mathbb{R}^s}(x)
\end{align*}

Split the above integral into two regions: $\Lambda = \{x \in \mathbb{R}^s: 0 \le \|x\|_1 \le \lambda|\log(\lambda^{s+1})| \}$ and $\Lambda^c$. Over $\Lambda^c$, $\exp(-||x||_1/\lambda) < \lambda^{s+1}.$
\begin{align*}
m_\lambda &= \int_{\Lambda^c}m(x) \exp\bigg(-\frac{||x||_1}{\lambda}\bigg)d\mu_{\mathbb{R}^s}(x) + \int_{\Lambda}m(x) \exp\bigg(-\frac{||x||_1}{\lambda}\bigg)d\mu_{\mathbb{R}^s}(x) \\
&=O\bigg( \lambda^{s+1} \bigg) + \int_{\Lambda} m(x) \bigg[1+O\bigg(\frac{1}{\lambda}\exp\bigg(-\frac{||x||_1}{\lambda}\bigg)\bigg)\bigg]d\mu_{\mathbb{R}^s}(x) \\
&=O\bigg( \lambda^{s+1} \bigg) +\int_{\Lambda} m(x) \bigg[1+O(\lambda^s) \bigg]d\mu_{\mathbb{R}^s}(x)
\end{align*}
Since $m(x)$ is continuous on an open neighborhood containing the origin, we may choose $\lambda$ small enough so that $m(x)$ is uniformly continuous on $\Lambda.$ Then, 
\begin{align*}
m_\lambda&= O\bigg( \lambda^{s+1} \bigg) + \int_{\Lambda} [m(0) + o(1)][1+O(\lambda^s)] d\mu_{\mathbb{R}^s} (x) 
= m(0) \frac{|2(s+1)\lambda \log \lambda |^s}{\Gamma(s+1)} + o(|\lambda \log \lambda|^s) 
\end{align*}
at leading order as $\lambda\to 0^+$. Here, $|2(s+1)\lambda \log \lambda|^s/\Gamma(s+1)$  is the Lebesgue measure of $\Lambda.$


We now turn to the small $\lambda$ behavior of $E_{\tilde{\Pi}}[g(\theta)].$  Similar to the study of $m_\lambda$, separate $E_{\tilde{\Pi}}[g(\theta)]$ into integrals over $\Lambda$ and $\Lambda^c$. Again, we may choose $\lambda$ sufficient small so that both $m(x)$ and \\ $
G(x)= \int_{\nu_{\mc D}^{(-1)}(x)} g(\theta) f(\theta)J^{-1}(\nu_{\mc D}(\theta)) d\bar{\mathcal{H}}^{r-s}(\theta) $
are continuous on $\Lambda$ and hence uniformly continuous at $x=0.$ Additionally, the positivity of $m(0)$ implies that $E[g(\theta)|\nu_{\mc D}(\theta)=x]$ is also uniformly continuous at $x=0.$ Therefore,
\begin{align*}
E_{\tilde{\Pi}}[g(\theta)] 
&=\int_{\Lambda^c} \frac{\exp\big(-\frac{||x||_1}{\lambda}\big)}{m_\lambda} m(x)E[g(\theta)|\nu_{\mc D}(\theta) = x] d\mu_{\mathbb{R}^s}(x) \\
&\hspace{0.4 cm}+ \int_{\Lambda} \frac{1+o(1)}{m(0)\frac{|2(s+1)\lambda \log \lambda|^s}{\Gamma(s+1)} +o(\lambda|\log \lambda|^s )} \big[m(0)+o(1)\big]\big[E[g(\theta)|\nu_{\mc D}(\theta)=0] + o(1) \big] d\mu_{\mathbb{R}^s}(x) \\
&= O\bigg(\frac{\lambda^{s+1}}{m_\lambda}\bigg) + E[g(\theta)|\nu_{\mc D}(\theta) = 0] + o(1) = E[g(\theta)|\theta \in \mathcal{D}] +O\bigg( \frac{\lambda}{|\log\lambda|^s}\bigg) + o(1).
\end{align*}
And we may conclude that $|E[g|\theta\in\mathcal{D}] - E_{\tilde{\Pi}}[g] | \to 0$  as $\lambda\to0^+.$

The proof of the final statement follows by changing the $o(1)$ correction within the integrals over $\Lambda$ to $O(\lambda |\log \lambda^{s+1}|)$ corrections.  As a result, the leading order error is then $O(\lambda|\log \lambda|^{-s})$ as $\lambda\to 0^+.$
\end{proof}

\newpage

{\huge \textbf{Supplementary Materials}}

\section*{S1: Constrained Subspaces with Positive Lebesgue Measure}
\subsection*{S1.1: Constructing $\|\nu_{\mc D}(\theta) \|$}
Unlike the measure zero constraints discussed in the article, positive measure constraints most often arise through inequalities involving one or more of the components of $\theta.$  Thus, a general choice of $\|\nu_{\mc D}(\theta)\|$ is a direct measure of the distance from $\theta$ to the closest point $\mc D.$ We focus on the simplest choice,
\begin{equation}
\|\nu^{(k)}_{\mc D}(\theta) \| = \inf_{x \in \mc D} \| \theta - x \|_k
\label{EQ:Direct_Distance}
\end{equation}
where $\| \cdot \|_k$ denotes the $k$-norm distance in $\mc R.$  

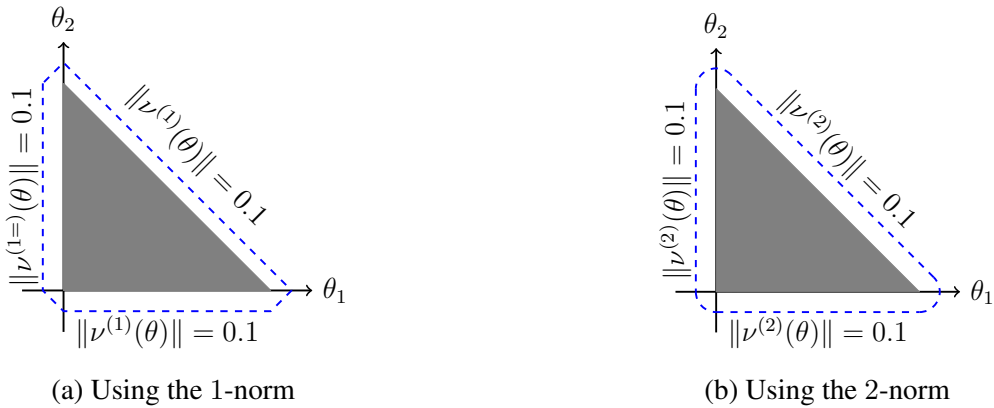
\begin{figure}[h]
\begin{subfigure}[b]{0.45\textwidth}
        \centering
        \resizebox{0.6\linewidth}{!}{
\begin{tikzpicture}[scale=3, line join=bevel]

\draw[help lines, color=gray!30, dashed] (-0.2,-0.2) grid (0.1,0.1);
\draw[->, thick] (-0.2,0)--(1.2,0) node[right]{$\theta_1$};
\draw[->, thick] (0,-0.2)--(0,1.2) node[above]{$\theta_2$};

\draw[gray, thin, fill =gray] (0,0) -- (1,0) -- (0,1) -- cycle;

\draw[blue, thick, dashed] (-0.1,1) -- (-0.1,0) --
 (-0.1,0) -- (0,-0.1)--
  (0,-0.1) -- (1,-0.1)-- 
  (1.1, 0) --  (0,1.1)  -- cycle;;


\node[rotate=90] at (-0.2,0.5) {$\| \nu^{(1=)}(\theta)\|=0.1$};
\node[rotate=0] at (0.5,-0.2) {$\| \nu^{(1)}(\theta)\|=0.1$};
\node[rotate=-45] at (0.65,0.65) {$\| \nu^{(1)}(\theta)\|=0.1$};

\end{tikzpicture}}
        \caption{Using the $1$-norm}
\end{subfigure}
\quad
    \begin{subfigure}[b]{0.45\textwidth}
        \centering
        \resizebox{0.6\linewidth}{!}{
\begin{tikzpicture}[scale=3, line join=bevel]

\draw[help lines, color=gray!30, dashed] (-0.2,-0.2) grid (0.1,0.1);
\draw[->, thick] (-0.2,0)--(1.2,0) node[right]{$\theta_1$};
\draw[->, thick] (0,-0.2)--(0,1.2) node[above]{$\theta_2$};

\draw[gray, thin, fill = gray] (0,0) -- (1,0) -- (0,1) -- cycle;
\draw[blue, thick, dashed] (-0.1,1) -- (-0.1,0);
\draw[blue, thick, dashed] (-0.1,0) arc (180:270:0.1);
\draw[blue, thick, dashed] (0,-0.1) -- (1,-0.1);
\draw[blue, thick, dashed] (1,-0.1) arc (-90:45:0.1);
\draw[blue, thick, dashed] (1.0707, 0.0707) --  (0.0707,1.0707); 
\draw[blue, thick, dashed] (0.0707,1.0707) arc (45:180:0.1);


\node[rotate=90] at (-0.2,0.5) {$\| \nu^{(2)}(\theta)\|=0.1$};
\node[rotate=0] at (0.5,-0.2) {$\| \nu^{(2)}(\theta)\|=0.1 $};
\node[rotate=-45] at (0.65,0.65) {$\| \nu^{(2)}(\theta)\|=0.1$};
\end{tikzpicture}
        }
        \caption{Using the $2$-norm}
\end{subfigure}

 \caption{The boundary of the neighborhood (blue dashed line) along $\|\nu^{(k)}_\mathcal{D}(\theta)\| = 0.1$  formed using two different choice of metric in \eqref{EQ:Direct_Distance}.}
 
\label{fig:two_distances}
\end{figure}

The choice of metric in \eqref{EQ:Direct_Distance} will effect the relaxation away from $\mc D.$  Similar to the measure zero case, one can compare the effects of this choice through the $d$-expansion of $\mc D.$  To illustrate, suppose $\theta \in \mathbb{R}^2$ and consider a  triangular constrained space arising from three inequalities $$\mc D=\{(\theta_1,\theta_2):  \theta_1 >0, \theta_2>0, \theta_1+\theta_2<1\}.$$ The level sets $\| \nu^{(k)}(\theta) \| = 0.1$, shown in Figure \ref{fig:two_distances} for $k=1,2$, are equivalent along the vertical and horizontal sections but differ along the corners and diagonal portion of the curves.

Similar to the measure zero case, the metric chosen in \eqref{EQ:Direct_Distance} should be based on prior belief about how the probability should decrease outside of D.  While the $k$-norms are natural options, particularly when the level curves of $\| \nu_{\mc D}(\theta)\|$ are similar to the boundary of $\mc D$, one may also elect for anistrophic relaxation if there is prior justifications for the choice.  Again, one can try several different choices, while assessing sensitivity of the results. 
\subsection*{S1.2: Constructing the Relaxed Posterior}
Once a choice of $\|\nu_{\mc D}(\theta)\|$ has been made the construction of the relaxed posterior is much simpler when $\mathcal{D}$ is a subset of $\mc R$ with positive Lebesgue measure.  The sharply constrained density is  a truncated version of the unconstrained one, with
$$\pi_\mathcal{D}(\theta\mid Y) = \frac{\mathcal{L}(\theta; Y)
\pi_\mathcal{R}(\theta)\mathbbm{1}_\mathcal{D}(\theta)}{\int_\mathcal{D}
\mathcal{L}(\theta; Y)
\pi_\mathcal{R}(\theta)d\mu_\mathcal{R}(\theta)}\propto \mathcal{L}(\theta;
Y) \pi_\mathcal{R}(\theta)\mathbbm{1}_\mathcal{D}(\theta), $$
which is defined with respect to $\mu_\mathcal{R}$.   For technical reasons, we consider only those cases where$\|\nu_{\mc D}(\theta) \| >0$ for $\mu_{\mc R}$-almost every $\theta \in \mc R \setminus \mc D.$ 
For constraint relaxation, we replace the indicator with an exponential function of distance
\begin{equation}
\label{EQ:relaxedDensityPosMeasure}
\tilde{\pi}_\lambda(\theta ) =
\frac{\mathcal{L}(\theta;
Y)\pi_\mathcal{R}(\theta)\exp\big(-
\lambda^{-1}{\|\nu_\mathcal{D}(\theta)\|}\big)}{\int_{\mathcal{R}}\mathcal{L}(\theta; Y)
\pi_\mathcal{R}(\theta)\exp\big(-{\lambda^{-1}}{\|\nu_\mathcal{D}(\theta)\|}\big)
d\mu_\mathcal{R}(\theta)} \propto
\mathcal{L}(\theta; Y)
\pi_\mathcal{R}(\theta)\exp\big(-{\lambda^{-1}}{\|\nu_\mathcal{D}(\theta)\|}\big)
\end{equation}
which is also absolutely continuous with respect to $\mu_\mathcal{R}.$  

Expression (\ref{EQ:relaxedDensityPosMeasure}) replaces the function $\mathbbm{1}_\mathcal{D}(\theta)$, which is equal to one for $\theta \in \mc D$ and zero for $\theta \not \in \mc D$, with 
$\exp\big(-{\lambda^{-1}}{\|\nu_\mathcal{D}(\theta)\|}\big)$, which is still equal to one for $\theta \in \mc D$ but decreases exponentially as $\theta$ moves away from $\mc D$.  The prior is effectively shrinking
$\theta$ towards $\mc D$, with the exponential tails reminiscent of the double exponential (Laplace) prior that forms the basis of the widely used Lasso procedure.  Potentially, we could allow a greater degree of robustness to the choice of $\mc D$ by choosing a heavier tailed function in place of the exponential; for example, using the kernel of a generalized double Pareto or t-density.  However, such choices introduce an additional hyperparameter, and we focus on the exponential for simplicity.  


As a simple illustrative example, we consider a Gaussian likelihood with inequality constraints on the mean.
In particular, let 
 $$y_i \stackrel{iid}{\sim} \No(\theta,1),\quad i =1,\ldots,n, \qquad \pi_\mathcal{R}(\theta) =\No(\theta; 0, 1000).$$ 
Suppose there is prior knowledge that $\theta<1$. The posterior under a sharply constrained model is
$$ \pi_{\mc D}(\theta \mid Y) \propto {\sigma^{-1}}\phi\bigg(\frac{\theta- \mu}{\sigma}\bigg) \mathbbm{1}_{\theta<1}, \quad \mu=    \frac{ \bar yn}{ 1/1000 + n },   \quad \sigma^2 = \frac{1}{ 1/1000 + n },$$
where $\phi$ denotes the density of the standard Gaussian. This posterior corresponds to 
$\No_{(-\infty,1)}(\mu,\sigma^2)$, which is a $\No(\mu,\sigma^2)$ distribution truncated to the region $\theta < 1$.

\begin{figure}[h]
\begin{subfigure}[b]{0.45\textwidth}
  \centering
 \includegraphics[width=0.6\textwidth]{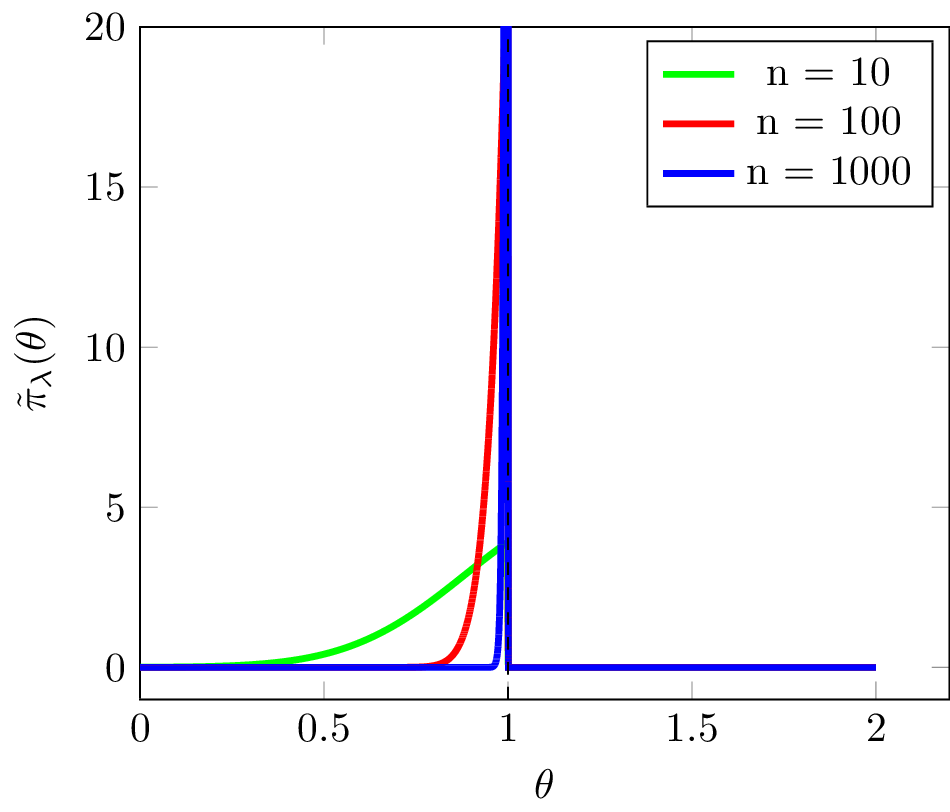}
 \caption{Sharply constrained posterior}
\end{subfigure}
\begin{subfigure}[b]{0.45\textwidth}
  \centering
 \includegraphics[width=0.6\textwidth]{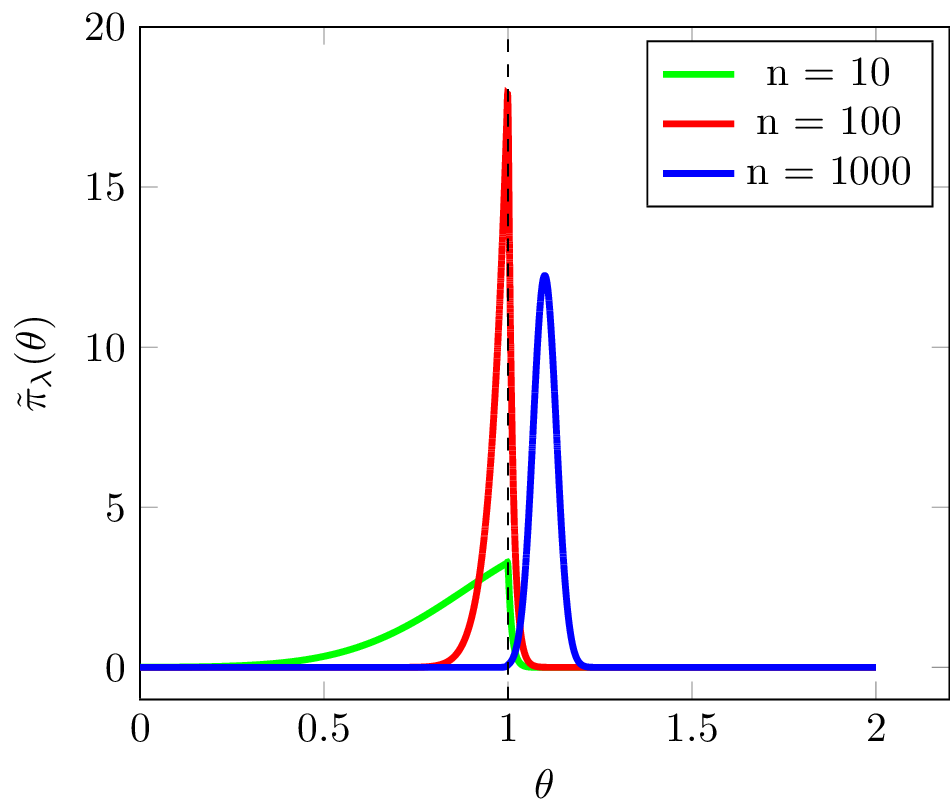}
 \caption{Constraint relaxed posterior}
\end{subfigure}
 \caption{Posterior densities for a Gaussian mean ($\theta$) under sharp (panel (a)) and relaxed constraints (panel (b)). The constraint region is $\mc D = (-\infty,1)$ and the true value is $\theta=1.2$, which falls slightly outside $\mc D$ representing misspecification.
\label{fig:gaussian_inequality}}
\end{figure}

If $\theta$ is indeed less than one, incorporation of the constraint has the benefit of reducing uncertainty in the posterior distribution, leading to greater concentration around the true value.  However, even slight mis-specification of the constrained region can lead to biased inferences; for example, perhaps $\theta=1.2.$
In this case, as the sample size $n$ increases, the sharply constrained posterior distribution $\No_{(-\infty,1)}(\mu,\sigma^2)$ becomes more and more concentrated near the $\theta=1$ boundary as illustrated in Figure~\ref{fig:gaussian_inequality}(a).

The constraint relaxed approach is well justified in this case as it allows some probability to be allocated to the $\theta>1$ region under the posterior: 
$$ \tilde{\pi}_{\lambda}(\theta ) \propto {\sigma^{-1}}\phi\bigg(\frac{\theta- \mu}{\sigma}\bigg) \exp\bigg(-\frac{(\theta-1)_+ }{\lambda}\bigg) , \quad \mu=    \frac{ \bar y n}{ 1/1000 + n },   \quad \sigma^2 = \frac{1}{ 1/1000 + n }$$
where $(\theta-1)_+$ is the direct distance to the constrained space. In this simple setting, our implementation of the constraint coincides with the approach proposed in Neal (2011). With a small value $\lambda=10^{-2}$, representing high prior concentration very close to $\mc D = (-\infty,1)$, the relaxed posterior is close to 
the sharply constrained one for small to moderate sample sizes, as illustrated in Figure~\ref{fig:gaussian_inequality}(b).  However, as $n$ increases, the posterior becomes concentrated around the true $\theta$ value, even when it falls outside of the constrained space.

\subsection*{S1.3: Theory} \label{SEC:Positive_measure_theory}


We focus on quantifying the difference between the sharply constrained and relaxed posterior distributions, 
both of which are absolutely continuous with respect to Lebesgue measure on $\mathcal{R}$.  The posterior expectation of $g$ under the sharply constrained prior is 
\begin{equation}
\label{EQ:Expectation_Positive_Measure_Constraint} 
E[g(\theta)\mid\theta\in\mathcal{D}] = \int_\mathcal{D}
g(\theta)\pi_\mathcal{D}(\theta \mid Y)d\mu_\mathcal{R}(\theta) =
\frac{\int_\mathcal{D} g(\theta)\mathcal{L}(\theta; Y)
\pi_\mathcal{R}(\theta)d\mu_\mathcal{R}(\theta)}{\int_\mathcal{D}
\mathcal{L}(\theta; Y)
\pi_\mathcal{R}(\theta)d\mu_\mathcal{R}(\theta)}.
\end{equation}
Similarly, the posterior expectation of $g$ under the relaxed prior is
\begin{equation} \label{EQ:Expectation_Positive_Measure_Relaxed}
E_{\tilde{\pi}_\lambda}[g(\theta)] = \int_\mathcal{R}
g(\theta)\tilde{\pi}_\lambda(\theta)d\mu_\mathcal{R}(\theta) =
\frac{\int_\mathcal{R} g(\theta)\mathcal{L}(\theta; Y)
\pi_\mathcal{R}(\theta)
\exp(-\|\nu_\mathcal{D}(\theta)\|/\lambda)d\mu_\mathcal{R}(\theta)}{\int_\mathcal{R}
\mathcal{L}(\theta; Y)
\pi_\mathcal{R}(\theta)\exp(-\| \nu_\mathcal{D}(\theta)\|/\lambda)d\mu_\mathcal{R}(\theta)}.\end{equation}
A short calculation shows that the magnitude of the difference $E[g(\theta)\mid\theta \in \mc D] - E_{\tilde{\pi}_{\lambda}}[g(\theta)]$ depends on two quantities: the posterior probability of $\mc D$ under the unconstrained posterior, 
and the average magnitude of $|g(\theta)|$ over $\mc R \setminus \mc D$ with respect to the relaxed posterior. These results are summarized in Lemma 1.

\begin{lemma} \label{THM:positive_measure_approximation_error} Suppose $g
\in \mathbb{L}^1(\mathcal{R},
\mathcal{L}(\theta;Y)\pi_\mathcal{R}(\theta)d\mu_\mathcal{R})$.  Then,
$$\Big|E[g(\theta) \mid \theta\in\mathcal{D}] -
E_{\tilde{\pi}_\lambda}[g(\theta)]   \Big| \le
\frac{\int_{\mathcal{R}\setminus \mathcal{D}}
(C_\mathcal{R}E|g(\theta)|+|g(\theta)|) \mathcal{L}(\theta; Y)
\pi_\mathcal{R}(\theta)\exp(-\lambda^{-1}\|\nu_\mathcal{D}(\theta)\| )
d\mu_\mathcal{R}(\theta)}{\big[\int_\mathcal{D} \mathcal{L}(\theta; Y)
\pi_\mathcal{R}(\theta)d\mu_\mathcal{R}(\theta)\big]^2 }$$ where
$E|g(\theta)| \propto \int_\mathcal{R} |g(\theta)|
\mathcal{L}(\theta;Y)\pi_\mathcal{R}(\theta) d\mu_\mathcal{R}(\theta)$ is the
expected value of $|g(\theta)|$ with respect to the unconstrained posterior
density and $C_\mathcal{R} = \int_\mathcal{R}
\mathcal{L}(\theta;Y)\pi_\mathcal{R}(\theta)d\mu_\mathcal{R}(\theta)$ is
the normalizing constant of this unconstrained posterior density.
Furthermore, if $\|\nu_\mathcal{D}(\theta)\|$ is 
positive for $\mu_{\mc R}$ almost every  $\theta\in \mathcal{R}\setminus\mathcal{D}$, it follows
from the dominated convergence theorem that $$\big| E[g(\theta)
\mid\theta\in\mathcal{D}] - E_{\tilde{\pi}_\lambda}[g(\theta)]  \big|\to 0
\text{ as } \lambda \to 0^+.$$ \end{lemma}
While this Lemma indicates $E_{\tilde{\pi}_\lambda}[g(\theta)] \to E[g(\theta)\mid\theta\in \mc D]$ as $\lambda \to 0^+$, for a fixed $\lambda>0$, large differences can arise if  (i) $|g(\theta)|\mathcal{L}(\theta;Y) \gg \exp(-\lambda^{-1} \|\nu_{\mc D}(\theta) \|)$ on average over a subset of $\mc R\setminus \mc D$ or (ii) the posterior probability of $\mc D$ is small with respect to the unconstrained posterior.  

With regards to (i), consider the case where $\mc F$ is a measurable subset of $\mc R$ for which $\mc F \cap \mc D = \emptyset$ and let $g(\theta) = \mathbbm{1}_{\mc F}(\theta)$. Then, $E[g(\theta)\mid\theta\in \mc D]=0.$  However,\ $E_{\tilde{\pi}_\lambda}[g(\theta)]$ may be large if $\mathcal{L}(\theta;Y) \gg \exp(-\lambda^{-1} \|\nu_{\mc D}(\theta) \|)$ for $\theta\in\mc F$.  As such, over $\mc F$ the likelihood is dominating the relaxation allowing the constraint relaxed density to assign positive probability to $\mc F$ which is not possible for the sharply constrained density.

In the case of (ii), the error is inversely proportional to the square of the unconstrained posterior probability of $\mc D.$ Thus, when $\theta \in \mc D$ is unlikely under the unconstrained model any relaxation away from the constraint is amplified.  This effect in particular demonstrates the usefulness of constraint relaxation. If constraints are misspecified and $\theta\in \mc D$ is not supported by the data, the posterior estimates using the relaxed density can display a large sensitivity to the choice of $\lambda$ indicating that the constraints themselves should be re-evaluated. 

Lemma 1 indicates that one can obtain sufficiently accurate estimates of
$E[g(\theta)\mid\theta\in\mathcal{D}]$ by sampling from $\tilde{\pi}_\lambda$ when
$\lambda$ is sufficiently small.  From a practical standpoint, it is
desirable to understand the rate at which
$E_{\tilde{\pi}_\lambda}[g(\theta)] $ converges to
$E[g(\theta)\mid\theta \in\mathcal{D}]$. The answer ultimately depends on the choice of distance function and its behavior on $\mc R\setminus \mc D.$ We supply the following theorem when the distance function $\|\nu_{\mc D}(\theta)\| = \inf_{x\in\mc D}\|\theta-x\|_2$. One can use the analysis contained in the proof of this theorem, contained in S2, as a guide to construct convergence rates for a different choice of $\|\nu_{\mc D}(\theta)\|.$ 

\begin{theorem} \label{THM:Positive_measure_convergence_rate} In addition to the assumptions of Lemma \ref{THM:positive_measure_approximation_error}, suppose $g
\in  \mathbb{L}^2(\mathcal{R},
\mathcal{L}(\theta;Y)\pi_\mathcal{R}(\theta)d\mu_\mathcal{R})$,
$\|\nu_\mathcal{D}(\theta)\|= \inf_{x\in\mathcal{D}} \|\theta-x\|_2$, $\mathcal{D}$ has a piecewise smooth boundary, and that
$\mathcal{L}(\theta;Y)\pi_\mathcal{R}(\theta)$ is continuous on an
open neighborhood containing $\mathcal{D}$.  Then for $0<\lambda
\ll 1,$ $$ \big|E[g(\theta) \mid\theta\in\mathcal{D}] -
E_{\tilde{\pi}_\lambda}[g(\theta)]   \big| = O(\lambda^{1/2}).  $$
\end{theorem} This theorem follows by applying the Cauchy-Schwartz
inequality to the term in the numerator of the bound given in Lemma
\ref{THM:positive_measure_approximation_error}.  This bound holds for general, even unbounded $\mc D$. More details regarding the coefficient in the error rate are contained in the proof but omitted here for brevity.


\section*{S2: Proofs for Main Results}
\label{APP:Positive_Measure_Convergence_Proofs}
\section*{S2.1: Lemma 1}
\begin{proof}Recall, that the distance function $\|\nu_\mathcal{D}(\theta)\|$ is chosen so that $\|\nu_\mathcal{D}(\theta)\|$ is zero for all $\theta\in \mathcal{D}$. It follows that for any function $g(\theta)$
\begin{equation}
\begin{split}
&\int_{\mathcal{R}}  g(\theta) \mathcal{L}(\theta;Y)\pi_\mathcal{R}(\theta)\exp(-\lambda^{-1}\|\nu_\mathcal{D}(\theta)\|) d\mu_\mathcal{R}(\theta) \\
&=  \int_{\mathcal{R}\setminus \mathcal{D}}g(\theta) \mathcal{L}(\theta;Y)\pi_\mathcal{R}(\theta)\exp(-\lambda^{-1}\|\nu_\mathcal{D}(\theta)\|) d\mu_\mathcal{R}(\theta) + \int_{ \mathcal{D}} g(\theta) \mathcal{L}(\theta;Y)\pi_\mathcal{R}(\theta) d\mu_\mathcal{R}(\theta) .
\end{split}
\label{EQ:Expectation_Identity_Positive_Measure}
\end{equation}

For brevity, we let $f(\theta) = \mathcal{L}(\theta;Y)\pi_\mathcal{R}(\theta)$ and use $df(\theta) = \mathcal{L}(\theta;Y)\pi_\mathcal{R}(\theta)d\mu_\mathcal{R}(\theta)$ throughout the proof. Then,
\begin{align*}
&\bigg| E[g(\theta)|\theta\in\mathcal{D}]- E_{\tilde{\pi}_\lambda}[g(\theta)]\bigg| 
= \bigg|\frac{ \int_\mathcal{D} g(\theta) df(\theta)}{\int_\mathcal{D} df(\theta)} - \frac{\int_{\mathcal{R}} g(\theta) \exp\big(-\lambda^{-1}\|\nu_\mathcal{D}(\theta)\|)df(\theta)}{\int_{\mathcal{R}}  \exp\big(-\lambda^{-1}\|\nu_\mathcal{D}(\theta)\|)df(\theta)} \bigg| \\
& = \bigg|\frac{\int_{\mathcal{R}\setminus \mathcal{D}} \exp(-\lambda^{-1}\|\nu_\mathcal{D}(\theta)\| ) df(\theta)  \int_\mathcal{D}g(\theta) df(\theta)-\int_\mathcal{D}df(\theta)  \int_{\mathcal{R}\setminus \mathcal{D} } g(\theta) \exp(-\lambda^{-1}\|\nu_\mathcal{D}(\theta)\|) df(\theta)}{\int_\mathcal{D} df(\theta)[\int_\mathcal{D} df(\theta) + \int_{\mathcal{R}\setminus\mathcal{D}}  \exp(-\lambda^{-1}\|\nu_\mathcal{D}(\theta)\|) df(\theta)] }  \bigg|
\end{align*}
where the second equality follows from combining the fractions and making use of \eqref{EQ:Expectation_Identity_Positive_Measure}. We can bound the denominator from below by  $ C_\mathcal{D}^2 = \big[\int_\mathcal{D} \mathcal{L}(\theta;Y)\pi_\mathcal{R}(\theta) d\mu_\mathcal{R}(\theta) \big]^2>0$ so that 
\begin{equation*}
\begin{split}
&\big| E[g(\theta)|\theta\in\mathcal{D}]-E_{\tilde{\pi}_\lambda}[g(\theta)]\big| \\ 
&\le \frac{\big|\int_{\mathcal{R}\setminus \mathcal{D}} \exp(-\lambda^{-1}\|\nu_\mathcal{D}(\theta)\| ) df(\theta)  \int_\mathcal{D}g(\theta) df(\theta) -\int_\mathcal{D}df(\theta)  \int_{\mathcal{R}\setminus \mathcal{D} } g(\theta)\exp(-\lambda^{-1}\|\nu_\mathcal{D}(\theta)\|) df(\theta)\big|}{C_\mathcal{D}^2 } 
\end{split}
\end{equation*} 
Add and subtract $$\int_{\mathcal{R}\setminus \mathcal{D}} \mathcal{L}(\theta;Y)\pi_\mathcal{R}(\theta)\exp(-\lambda^{-1}\|\nu_\mathcal{D}(\theta)\| ) d\mu_\mathcal{R}(\theta)  \int_{\mathcal{R}\setminus \mathcal{D}} g(\theta) \mathcal{L}(\theta;Y)\pi_\mathcal{R}(\theta)\exp(-\lambda^{-1}\|\nu_\mathcal{D}(\theta)\| ) d\mu_\mathcal{R}(\theta)  $$ within the numerator, and apply the triangle inequality. Thus,

\begin{align*}
&\big| E[g(\theta)|\theta\in\mathcal{D}]-E_{\tilde{\pi}_\lambda}[g(\theta)]\big| \\
& \le \frac{ \bigg| \int_{\mathcal{R}\setminus \mathcal{D}} \exp(-\lambda^{-1}\|\nu_\mathcal{D}(\theta)\| ) df(\theta) \bigg|  \bigg|\int_\mathcal{D}g(\theta)df(\theta) - \int_{\mathcal{R}\setminus \mathcal{D}} g(\theta)\exp(-\lambda^{-1}\|\nu_\mathcal{D}(\theta)\| ) df(\theta) \bigg|}{C_\mathcal{D}^2 }\\
& \hspace{2cm} + \frac{\bigg| \int_{\mathcal{R}\setminus \mathcal{D}} g(\theta)\exp(-\lambda^{-1}\|\nu_\mathcal{D}(\theta)\| ) df(\theta) \bigg|  \bigg|\int_\mathcal{D} df(\theta)- \int_{\mathcal{R}\setminus \mathcal{D}}  \exp(-\lambda^{-1}\|\nu_\mathcal{D}(\theta)\| )df(\theta) \bigg|}{C_\mathcal{D}^2 }
\end{align*}
Since $g\in\mathbb{L}^1(\mathcal{R},\mathcal{L}(\theta;Y)\pi_\mathcal{R}(\theta)d\mu_\mathcal{R})$, we can bound the numerators. First,
\begin{align*}
&\bigg| \int_{\mathcal{R}\setminus \mathcal{D}} \exp(-\lambda^{-1}\|\nu_\mathcal{D}(\theta)\| ) df(\theta) \bigg|  \bigg|\int_\mathcal{D} g(\theta) df(\theta) - \int_{\mathcal{R}\setminus \mathcal{D}} g(\theta) \exp(-\lambda^{-1}\|\nu_\mathcal{D}(\theta)\| )df(\theta) \bigg| \\
&\le \int_{\mathcal{R}\setminus \mathcal{D}} \exp(-\lambda^{-1}\|\nu_\mathcal{D}(\theta)\| ) df(\theta)    \int_{\mathcal{R}} |g(\theta)| df(\theta)   = C_\mathcal{R} E|g(\theta)| \int_{\mathcal{R}\setminus \mathcal{D}}\exp(-\lambda^{-1}\|\nu_\mathcal{D}(\theta)\| ) df(\theta).
\end{align*}
Here, $C_\mathcal{R} = \int_\mathcal{R}df(\theta)$ is the normalizing constant of $\mathcal{L}(\theta;Y)\pi_\mathcal{R}(\theta).$
Secondly,
\begin{align*}
&\bigg| \int_{\mathcal{R}\setminus \mathcal{D}} g(\theta) \exp(-\lambda^{-1}\|\nu_\mathcal{D}(\theta)\| )df(\theta) \bigg|  \bigg|\int_\mathcal{D} df(\theta) - \int_{\mathcal{R}\setminus \mathcal{D}} \exp(-\lambda^{-1}\|\nu_\mathcal{D}(\theta)\| ) df(\theta)  \bigg| \\
& = C_\mathcal{R}\int_{\mathcal{R}\setminus \mathcal{D}} |g(\theta)| \exp(-\lambda^{-1}\|\nu_\mathcal{D}(\theta)\| ) df(\theta).
\end{align*}
Thus, we have the bounds specified by the theorem,
\begin{align*}
&\big| E[g(\theta)|\theta\in\mathcal{D}]-E_{\tilde{\pi}_\lambda}[g(\theta)]\big| 
= \frac{C_\mathcal{R}\int_{\mathcal{R}\setminus \mathcal{D}} (E|g(\theta)|+|g(\theta)|) )\exp(-\lambda^{-1}\|\nu_\mathcal{D}(\theta)\| ) df(\theta)}{C_\mathcal{D}^2 }.
\end{align*}

By assumption, $g\in\mathbb{L}^1(\mathcal{R},\mathcal{L}(\theta;Y)\pi_\mathcal{R}(\theta)d\mu_\mathcal{R})$ and $\|\nu_\mathcal{D}(\theta)\| >0$ for $\mu_\mathcal{R}$ almost every   $\theta \in \mathcal{R}\setminus \mathcal{D}$. It follows that $( E|g(\theta)|+|g(\theta)|) f(\theta)$ is a dominating function of $( E|g(\theta)|+|g(\theta)|) f(\theta)\exp(-\lambda^{-1}\|\nu_\mathcal{D}(\theta)\| )$ which converges to zero for $\mu_\mathcal{R}$ almost every  $\theta\in\mathcal{R}\setminus\mathcal{D}$ as $\lambda\to 0^+.$ Thus, $\big| E[g(\theta)|\theta\in\mathcal{D}]-E_{\tilde{\pi}_\lambda}[g(\theta)]\big|\to 0$ as $\lambda\to0^+.$

\end{proof}

\subsection*{S2.2: Theorem 1}
\begin{proof}
We begin with the bound from Lemma 1. 
$$\big| E[g(\theta)|\theta\in\mathcal{D}]-E_{\tilde{\pi}_\lambda}[g(\theta)]\big| \le \frac{ C_\mathcal{R}\int_{\mathcal{R}\setminus \mathcal{D}} (E|g(\theta)|+|g(\theta)|) \mathcal{L}(\theta;Y)\pi_\mathcal{R}(\theta)\exp(-\lambda^{-1}\|\nu_\mathcal{D}(\theta)\| ) d\mu_\mathcal{R}(\theta)}{C_\mathcal{D}^2 }.$$
Applying the Cauchy-Schwartz inequality to the numerator,
\begin{align*}
&C_\mathcal{R} \int_{\mathcal{R}\setminus \mathcal{D}} (E|g(\theta)|+|g(\theta)|)\exp(-\lambda^{-1}\|\nu_\mathcal{D}(\theta)\| )df(\theta) \\
&\le C_\mathcal{R}\bigg(\int_{\mathcal{R}\setminus \mathcal{D}}   (E|g(\theta)|+|g(\theta)|)^2 df(\theta)\bigg)^{1/2} \bigg(\int_{\mathcal{R}\setminus \mathcal{D}}\exp(-2\lambda^{-1}\|\nu_\mathcal{D}(\theta)\| )df(\theta)\bigg)^{1/2} 
\end{align*}
By assumption, $g\in\mathbb{L}^2(\mathcal{R},\mathcal{L}(\theta;Y)\pi_\mathcal{R}(\theta)d\mu_\mathcal{R}).$ Thus,
\begin{align*}
&C_\mathcal{R} \int_{\mathcal{R}\setminus \mathcal{D}} (E|g(\theta)|+|g(\theta)|)\exp(-\lambda^{-1}\|\nu_\mathcal{D}(\theta)\| )df(\theta) \le C_{g}\bigg(\int_{\mathcal{R}\setminus \mathcal{D}}\exp(-2\lambda^{-1}\|\nu_\mathcal{D}(\theta)\| )df(\theta)\bigg)^{1/2}
\end{align*}
where $C_g=\bigg(3C_\mathcal{R}^2(E|g|)^2 + C_\mathcal{R}E[|g|^2] \bigg)^{1/2}.$ 
We separate the integral $\int_{\mathcal{R}\setminus \mathcal{D}}\exp(-2\lambda^{-1}\|\nu_\mathcal{D}(\theta)\| )df(\theta)$
over the sets $\Lambda=\{\theta: \, 0< \|\nu_\mathcal{D}(\theta)\|< -\lambda\log\lambda\}$ and $\Lambda^c=\{\theta: \|\nu_\mathcal{D}(\theta)\|> -\lambda\log\lambda\}$.
\begin{align*}
&\int_{\mathcal{R}\setminus \mathcal{D}}\exp(-2\lambda^{-1}\|\nu_\mathcal{D}(\theta)\| )df(\theta) 
= \int_{\Lambda^c}\exp(-2\lambda^{-1}\|\nu_\mathcal{D}(\theta)\| )df(\theta) +\int_{\Lambda}\exp(-2\lambda^{-1}\|\nu_\mathcal{D}(\theta)\| )df(\theta)\\
&\le \lambda^2 \int_{\Lambda^c}df(\theta) +\int_{\Lambda}\exp(-2\lambda^{-1}\|\nu_\mathcal{D}(\theta)\| ) df(\theta)=\ C_\mathcal{R}\lambda^2 +\int_{\Lambda}\exp(-2\lambda^{-1}\|\nu_\mathcal{D}(\theta)\| )df(\theta)
\end{align*}
From the requirements of Theorem 1 we now let $\|\nu_\mathcal{D}(\theta)\| = \inf_{x\in \mathcal{D}}||\theta - x||_2$ and assume that $\mathcal{D}$ has a piecewise smooth boundary.  In this case, the set $\Lambda=\{\theta: \,0 < \|\nu_\mathcal{D}(\theta)\|< -\lambda\log\lambda\}$ forms a `shell' of thickness $-\lambda \log \lambda$ which encases $\mathcal{D}.$ 


In this case, $J(\nu_\mathcal{D}(\theta))= 2.$ By the co-area formula,
$$\int\limits_{\Lambda}\exp(-2\lambda^{-1}\|\nu_\mathcal{D}(\theta)\| )df(\theta)  = \int_0^{-\lambda\log\lambda} e^{-\frac{x}{\lambda}}\bigg( \int_{\nu_\mathcal{D}^{-1}(x)} \frac{1}{2}f(\theta) d\bar{\mathcal{H}}^{r-1}(\theta)\bigg) dx$$
Again, we may take $\lambda$ sufficiently small so that $f(\theta)=\mathcal{L}(\theta;Y) \pi_\mathcal{R}(\theta) $ is continuous on $\Lambda.$  As such, the function $ \int_{\nu_\mathcal{D}^{-1}(x)} \frac{1}{2}f(\theta) d\bar{\mathcal{H}}^{r-1}(\theta)$ is a continuous map from the closed interval $[0,-\lambda\log\lambda]$ to $\mathbb{R}.$; hence, it is bounded.  As a result, 
\begin{align*}
&\int\limits_{\Lambda}\exp(-2\lambda^{-1}\|\nu_\mathcal{D}(\theta)\| )df(\theta) \le \sup_{x \in [0,-\lambda\log\lambda]} \bigg( \int_{\nu_\mathcal{D}^{-1}(x)} \frac{1}{2}\mathcal{L}(\theta;Y) \pi_\mathcal{R}(\theta) d\bar{\mathcal{H}}^{r-1}(\theta)\bigg)  \int_0^{-\lambda\log \lambda} e^{-\frac{x}{\lambda}}dx \\
& = \sup_{x \in [0,-\lambda\log\lambda]} \bigg( \int_{\nu_\mathcal{D}^{-1}(x)} \frac{1}{2}\mathcal{L}(\theta;Y) \pi_\mathcal{R}(\theta) d\bar{\mathcal{H}}^{r-1}(\theta)\bigg)   (\lambda - \lambda^2) = O(\lambda)
\end{align*}
Thus, we may conclude that
\begin{align*}
&\big| E[g(\theta)|\theta\in\mathcal{D}]-E_{\tilde{\pi}_\lambda}[g(\theta)]\big| \le  \frac{C_{g}}{C_\mathcal{D}^2}\bigg(C_\mathcal{R}\lambda^2 + \sup_{x \in [0,-\lambda\log\lambda]} \bigg( \int_{\nu_\mathcal{D}^{-1}(x)} \frac{1}{2}f(\theta) d\bar{\mathcal{H}}^{r-1}(\theta)\bigg)   (\lambda - \lambda^2) \bigg)^{1/2} \\
&= \frac{C_{g}}{C_\mathcal{D}^2}  \sup_{x \in [0,-\lambda\log\lambda]} \bigg( \int_{\nu_\mathcal{D}^{-1}(x)} \frac{1}{2}f(\theta)d\bar{\mathcal{H}}^{r-1}(\theta)\bigg)  \lambda^{1/2} + o(\lambda^{1/2})
\end{align*}
Since $\sup_{x \in [0,-\lambda\log\lambda]} \bigg( \int_{\nu_\mathcal{D}^{-1}(x)} \frac{1}{2}f(\theta) d\bar{\mathcal{H}}^{r-1}(\theta)\bigg)$ is a decreasing function in $\lambda$, it follows that $$\big| E[g(\theta)|\theta\in\mathcal{D}]-E_{\tilde{\pi}_\lambda}[g(\theta)]\big| = O(\lambda^{1/2})
$$ as $\lambda \to 0^+.$
\end{proof}

\section*{S3: Computing Efficiency in Constraint Relaxed Hamiltonian Monte Carlo}
\label{SEC:Efficiency}

It is interesting to study the effect of relaxation on computing efficiency of Hamiltonian Monte Carlo.
In understanding computational efficiency of Hamiltonian Monte Carlo, it is useful to
consider the number of leapfrog steps to be a function of $\dt$ and set $L
= \lfloor \tau / \dt \rfloor$ for a fixed integration time $\tau > 0$. In
this case, the mixing rate is determined by $\tau$ in the
limit $\dt \to 0$ \citep{betancourt17}. While a smaller
stepsize $\dt$ leads to a more accurate numerical approximation of
Hamiltonian dynamics and hence a higher acceptance rate, it takes a larger
number of leapfrog steps and gradient evaluations to achieve good mixing.
For computational efficiency, the stepsize $\dt$ should be chosen only as 
small as needed to achieve a reasonable acceptance rate \citep{beskos13, 
betancourt14}. A critical factor in determining a reasonable stepsize is 
the \textit{stability limit} of the leapfrog integrator \citep{neal2011mcmc}. 
When $\dt$ exceeds this limit, the approximation becomes unstable and the 
acceptance rate drops dramatically. Below the stability limit, the acceptance rate 
$a(\dt)$ of Hamiltonian Monte Carlo increases to 1 quite rapidly as $\dt \to 0$ and satisfies
$a(\dt) = 1 - \mc O(\dt^4)$ \citep{beskos13}.

For simplicity, the following discussions assume the mass matrix $\mass$ is
the identity, and $\mc D= \cap_{j=1}^s\{\theta :\nu_j(\theta)=0 \}$.
We denote $\mc D_j= \{\theta :\nu_j(\theta)=0 \}$ and consider a directional
relaxation, which lets\\
 $\exp(-\sum_j{\|\nu_j(\theta^*)\|}{\lambda_j^{-1}})$. 
Typically, the stability limit of the leapfrog integrator is closely related to the largest
eigenvalue $\xi_1(\theta)$ of the Hessian matrix $\hess_U(\theta)$ of
$U(\theta) = - \log \pi(\theta)$. Linear stability analysis and
empirical evidence suggest that, for stable approximation of Hamiltonian
dynamics by the leapfrog integrator in $\bb R^p$, the condition $\dt <
2\xi_1(\theta)^{-1/2}$ must hold on most regions of the parameter space
\citep{hairer06}. Under the Constraint Relaxation framework, the Hessian is given by 
\begin{equation} \label{eq:hessian_extrinsic}
\hess_U(\theta) = -\hess_{\log
(\mathcal L(\theta;y) \pi_{\mc R}(\theta))
}(\theta)+\sum_j \lambda_j^{-1} \hess {\|\nu_j(\theta)\|}
\mathbbm{1}_{\theta\not\in \mc D_j}. \end{equation}
For ${\theta\not\in \mc D_j}$, as we make relaxations tighter i.e.\ $\lambda^{}_j \to 0$, the second term dominates the eigenvalue in the first term and the largest eigenvalue effectively becomes
proportional to $ \underset{j: \theta \not\in \mc D_j}{\min}\lambda_j^{-1}$. In other words, if we think of the Hessian as representing local covariance structure in the target distribution, then the effect of constraints on the stability limit becomes significant roughly speaking when $\min_j \lambda_j^{-1}$ is chosen smaller than the variance of the distribution along ${\mc D}$. 

The above discussion shows that a choice of extremely small $\lambda_j$ --- corresponding to very tight constraints --- could create a computational bottleneck for Hamiltonian Monte Carlo. Additionally, very tight constraints make it difficult for the no-U-turn criterion of \cite{hoffman2014no} to appropriately calibrate the number of leapfrog steps because the U-turn condition may be met too early to adequately explore the parameter space. For this reason, it is in general best not to make constraints tighter than necessary. On the other hand, when the leapfrog integrator requires a stepsize $\epsilon \ll \min_j \lambda_j^{-1 / 2}$ for an accurate approximation, one can safely make the constraint tighter as desired without affecting computational efficiency of Hamiltonian Monte Carlo. 

In our experience, a small number of experiments with different values of $\lambda$'s were sufficient to find out when the constraint starts to become a bottleneck. Also, Hamiltonian Monte Carlo usually achieved satisfactory sampling efficiency under reasonably tight constraints. We now use a problem of sampling from the von Mises--Fisher distribution to illustrate how a choice of $\lambda$ affects sampling efficiency.

We test $\lambda = 10^{-3}$, $10^{-4}$ and $10^{-5}$ for Constraint Relaxed Hamiltonian Monte Carlo. Table~\ref{table_circle} shows the effective sample size per $1000$ iterations, the effective `violation' of the constraint $\|\nu^{(2)}(\theta)\|=|\theta_1^{2}+\theta_2^{2}-1|$, and the difference of the quantity $$\bigg |E_\Pi[\sum_j\theta_j] - E_{\tilde{\Pi}}[\sum_j\theta_j] \bigg |$$ as the measure of relaxation. As the expectation is numerically computed, to provide a baseline, we also compare two independent samples from the same exact distribution. The expectation difference based on $\lambda= 10^{-5}$ is indistinguishable from this low difference, while the other $\lambda$  have slightly larger expectation difference but more effective samples. 

\begin{table}[h]
\begin{center}

\begin{tabular}{ c| c | c| c  | c}  
\cline{2-4} 
& \multicolumn{3}{c|}{ Constraint Relaxed Hamiltonian Monte Carlo}   &     \\       
\cline{2-5}         
&  $\lambda=1\e\text{-}3$ & $\lambda=1\e\text{-}4$ & $\lambda=1\e\text{-}5$ &  Exact
\\
\hline
\hline
$\bigg |E_\Pi[\sum_j\theta_j] - E_{\tilde{\Pi}}[\sum_j\theta_j] \bigg |$ &  $2.5\e\text{-}2$ & $1.6\e\text{-}2$ &  $8\e\text{-}3$  &  $9\e\text{-}3$\\

&  $(1.4\e\text{-}2,6.5\e\text{-}2$) &$(1.2\e\text{-}2, 1.9\e\text{-}2)$ &  $(6\e\text{-}3,1.5\e\text{-}2)$   & $(7\e\text{-}3,1.5\e\text{-}2)$\\

\hline
$|\nu^{(2)}_\mathcal{D}(\theta)|$ 
& $9\e\text{-}4 $ 
& $9\e\text{-}5 $ 
& $9\e\text{-}6 $  & 0\\
& $(2.6 \e\text{-}5, 3.3\e \text{-}3)$& $(2.0\e \text{-}6, 3.4\e\text{-}4)$& $(2.7 \e\text{-}7, 3.5\e\text{-}5)$&   \\
\hline
ESS /1000 Iterations &  751 & 261 & 57 & 788    \\
\hline  
\end{tabular}
\end{center}
\caption{Benchmark of constraint relaxation methods on sampling von--Mises Fisher distribution on a unit circle. For each constraint relaxed posterior, the average expectation difference  (with 95\% credible interval, out of $10$ repeated experiments) is computed, and numeric difference is shown under column `exact' as comparing two independent copies from the exact distribution\label{table_circle}.
Effective sample size shows constraint relaxation with relatively large $\lambda$ has high computing efficiency.
}
\end{table}

\section*{S4: Support Expansion Near a Curved Torus}

Let $\mc R = \mathbb{R}^3$ and consider a curved torus
$$\mc D= \big\{\theta: (\theta_1,\theta_2,\theta_3) = \big(
(1+ 0.5 \cos\alpha_1) \cos\alpha_2, (1+ 0.5\cos\alpha_1) \sin\alpha_2,0.5 \sin\alpha_1\big), (\alpha_1,\alpha_2)\in [0,2\pi)^2 \big\},$$ 
which has intrinsic dimension two and zero three-dimensional Lebesgue measure, $\mu_{\mc R}(\mc D)=0$. \cite{diaconis2013manifold} proposed an algorithm for sampling from a uniform density with respect to Hausdorff measure over this compact manifold.

\begin{figure}[h]
\begin{subfigure}[b]{1\textwidth}
  \centering
 \includegraphics[width=0.6\textwidth]{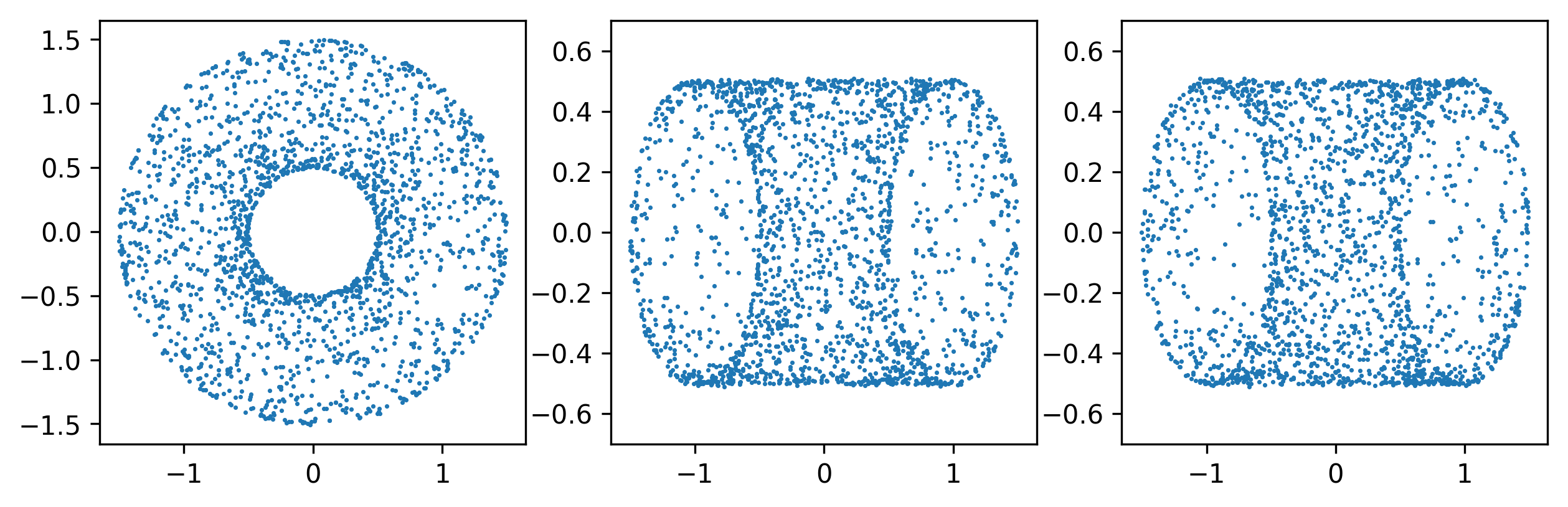}
 \caption{Constraint relaxed density with $\lambda = 0.01$}
\end{subfigure}
\begin{subfigure}[b]{1\textwidth}
  \centering
 \includegraphics[width=0.6\textwidth]{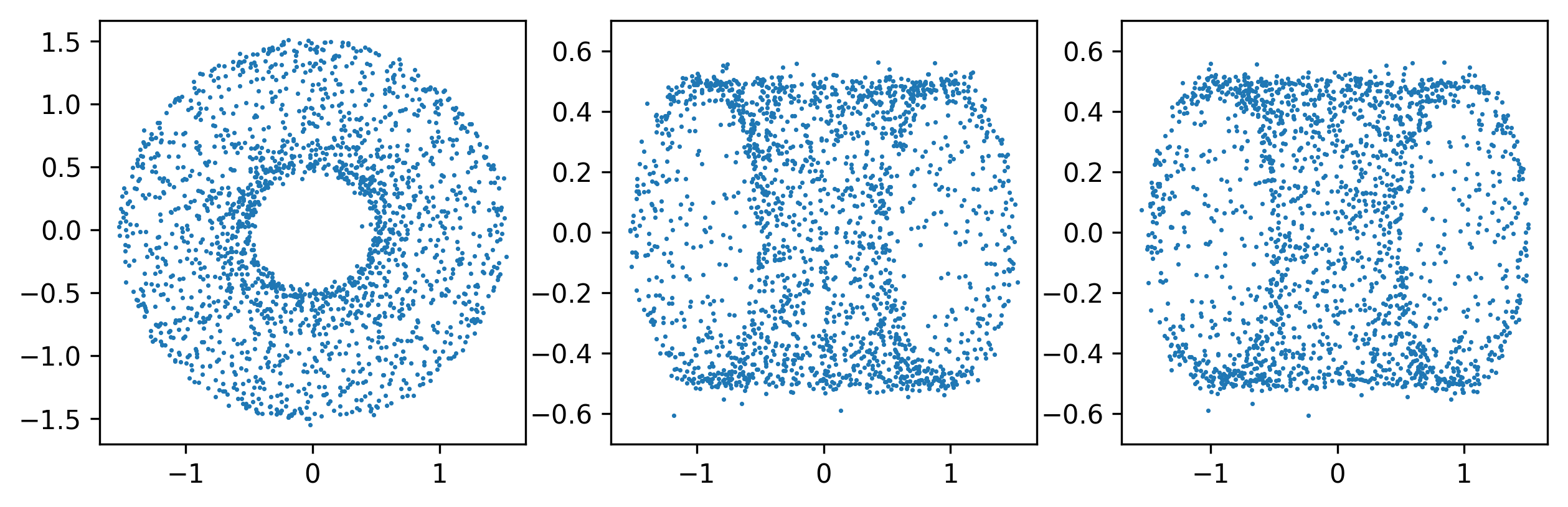}
 \caption{Constraint relaxed density  with $\lambda = 0.1$}
\end{subfigure}
 \caption{Samples from constraint relaxed based on a uniform density on a torus.  As $\lambda$ increases, more points are generated outside of the torus.
 \label{fig:torus}}
\end{figure}

The torus $\mc D$ can be defined implicitly as the solution set to the equation
$$\nu_{\mc D}(\theta) = \Big(1-\big(\theta_1^2+\theta_2^2\big)^{1/2}\Big)^2 + \theta_3^2 - \frac{1}{4} = 0.$$ 
Using this, we can replace the uniform density over the torus with the relaxed density
\begin{equation}
\begin{aligned}
\tilde\pi_{\lambda}(\theta)& \propto J(\nu_{\mc D}(\theta)) \exp(-\lambda^{-1}\|\nu_\mathcal{D}(\theta)\|)\\
& = 2\bigg[\bigg(1 - \big(\theta_1^2+\theta_2^2\big)^{1/2}\bigg)^2 + \theta_3^2\bigg]^{1/2} \exp\bigg\{-\lambda^{-1}\bigg| \Big(1-\big(\theta_1^2+\theta_2^2\big)^{1/2}\Big)^2 + \theta_3^2 - \frac{1}{4} \bigg| \bigg\}
\end{aligned}
\label{EQ:torus_relaxed_dens}
\end{equation}
which is defined with respect to $3$-dimensional Lebesgue measure. 
Here we initially multiplied the relaxed distribution by the Jacobian so that we attain uniform sampling on the torus under the sharp constraint, 
$$\pi_{\mc D}(\theta) \propto \frac{J(\nu_{\mc D}(\theta) )\mathbbm{1}_{\mc
    D}(\theta)}{J(\nu_{\mc D}(\theta))} = \mathbbm{1}_{\mc D}(\theta).$$
Figure~\ref{fig:torus} plots random samples from relaxed distribution to
uniform densities over the torus for two different values of $\lambda$,
corresponding to different degrees of relaxation.

%

\end{document}